\documentclass[10pt]{article}
\usepackage{a4wide}
\usepackage{graphicx}
\usepackage{amssymb}
\usepackage{epsfig, subfigure}
\usepackage{epstopdf}
\usepackage{algorithm}
\usepackage{algorithmic}
\usepackage{amsmath}
\usepackage{amsfonts}
\usepackage{url}
\usepackage{chemarr}
\usepackage{bbold}
\usepackage[version=3]{mhchem}

\newtheorem{example}{Example}

\newtheorem{remark}{Remark}
\newtheorem{proposition}{Proposition} 

\newcommand{\dd}{\mbox{$\mathbb{d}$}}
\newcommand{\ddone}{\mbox{$\breve{\mathbb{d}}$}}

\begin{document} 
\title{Extended master equation models for molecular communication networks} 
\author{Chun Tung Chou \\
School of Computer Science and Engineering, \\ University of New South Wales, \\ Sydney, Australia. 
E-mail: ctchou@cse.unsw.edu.au}

%

\maketitle

\begin{abstract}
We consider molecular communication networks consisting of transmitters and receivers distributed in a fluidic medium. In such networks, a transmitter sends one or more signalling molecules, which are diffused over the medium, to the receiver to realise the communication. In order to be able to engineer synthetic molecular communication networks, mathematical models for these networks are required. This paper proposes a new stochastic model for molecular communication networks called reaction-diffusion master equation with exogenous input (RDMEX). The key idea behind RDMEX is to model the transmitters as time series of signalling molecule counts, while diffusion in the medium and chemical reactions at the receivers are modelled as Markov processes using master equation. An advantage of RDMEX is that it can readily be used to model molecular communication networks with multiple transmitters and receivers. For the case where the reaction kinetics at the receivers is linear, we show how RDMEX can be used to determine the mean and covariance of the receiver output signals, and derive closed-form expressions for the mean receiver output signal of the RDMEX model. These closed-form expressions reveal that the output signal of a receiver can be affected by the presence of other receivers. 
Numerical examples are provided to demonstrate the properties of the model. 
\end{abstract} 


\section{Introduction} 
We consider molecular communication networks consisting of transmitters and receivers distributed in a fluidic medium. In such networks, a transmitter sends one or more signalling molecules, which are diffused over the medium, to the receiver to realise the communication. The study of molecular communication has its origin in biology and biophysics. Molecular communication is a vital mechanism in multi-cellular organisms. The human body, which has an estimated $10^{14}$ cells, uses molecular communication to keep the body in a healthy state. In fact, cells in the human body constantly communicate with other cells using molecular communication.

There are a couple of reasons why synthetic molecular communication networks, which are inspired by molecular communication in living organisms, should be studied \cite{Akyildiz:2008vt,Moore:2009eu,Hiyama:2010jf}. Firstly, synthetic molecular communication networks can be combined with nano-sensors and molecular computing \cite{Xie:2011fi} to form nano-sensor networks \cite{Atakan:2012ej} for health monitoring, medical diagnosis and cancer therapy. Secondly, the study of synthetic molecular communication networks can be used to enhance our understanding of their biological counterparts.

In order to be able to engineer synthetic molecular communication networks, we need mathematical models which can be used to predict the performance of these networks. For example, if a transmitter in a molecular communication network emits a number of signalling molecules to communicate with a receiver, we would like to be able to determine the receiver output signal in order to determine the probability of correct reception at the receiver. Such evaluations can be realised if a mathematical model is available to determine receiver output signal based on the transmitter's input signal. The main contribution of this paper is that we propose a new stochastic model for molecular communication networks. Our model is based on reaction-diffusion master equation (RDME) \cite{Gardiner} which is a well known model in physics and chemistry for modelling systems with both diffusion and chemical reactions. In this paper, we propose an extension to RDME, which we call reaction-diffusion master equation with exogenous input (RDMEX). The key idea behind RDMEX is to model the transmitters as time series of signalling molecule counts, while diffusion in the medium and chemical reactions at the receivers are modelled by RDME. An advantage of RDMEX is that it can readily be used to model  molecular communication networks with many transmitters and many receivers. For the case where the reaction kinetics at the receivers is linear, we show how RDMEX can be used to determine the mean and covariance of receiver output signal of molecular communication networks. These results allow us to derive closed-form expressions showing how the receiver outputs relate to the transmitter signals when there are multiple transmitters and receivers. These expressions show that the output of a receiver can be influenced by the presence of other receivers in a molecular communication network. They also reveal the coupling between diffusion and chemical reactions at the receivers. 

This paper is organised as follows. In Section \ref{sec:me}, we present background materials on master equations. In Section \ref{sec:rdmex_main}, we present the RDMEX and show how it can be used to model molecular communication networks with multiple transmitters and receivers. We also show in this section, for the case where the reaction kinetics at receivers is linear,  how we can determine the mean and covariance of the receiver output signals in molecular communication networks. The rest of the paper is focused on determining the mean receiver output signal. We approach this by using two methods, which will be discussed in sections \ref{sec:cont} and \ref{sec:zt}. In Section \ref{sec:cont}, we determine the continuum limit (i.e. infinite spatial resolution) of the RDMEX and show that it results in a reaction-diffusion partial differential equation (RDPDE). We derive a closed-form solution to this RDPDE and interpret the results. In Section \ref{sec:zt}, we determine the mean receiver output signal of RDMEX with finite spatial resolution and derive another closed-form solution. Numerical results are then presented in section \ref{sec:num} to show the accuracy of our solutions. Finally, section \ref{sec:related} describes the related work and section \ref{sec:con} gives the conclusions.

\section{Background on master equations} 
\label{sec:me} 
The aim of this section is to provide the necessary background on master equations. The treatment here is brief and includes only the results needed for this article. The reader can refer to the texts \cite{Gardiner,vanKampen} or tutorial article \cite{Erban:2007we} for a more complete treatment of this subject. This section is divided into two parts. We first introduce the general master equation and give a simple example on how to use master equation to model a chemical reaction. We then quote some results on mean and covariance of the Markov processes modelled by master equations. 

\subsection{General master equation}
Consider a continuous-time integer-value vector Markov process $Q(t) \in \mathbb{Z}^p$, where $p$ is the number of vector components, $\mathbb{Z}$ is the set of all integers and $t$ is time. When the Markov process $Q(t)$ is in state $q \in \mathbb{Z}^p$, it jumps to the state $q + r_j$ (where $r_j \in \mathbb{Z}^p$ with $j = 1,2,...J$ and $J$ is the total number of possible jumps) at a transition rate of $W_j(q)$. Let $P(q,t | q_0, t_0)$ denote the conditional probability that $Q(t) = q$ given that $Q(t_0) = q_0$. 

We are interested to determine how $P(q,t | q_0, t_0)$ evolves over time. We can do this by using a coupled set of ordinary differential equations (ODEs) known as the master equation: 
\begin{eqnarray}
\frac{d P(q,t | q_0, t_0)}{dt} & = & \sum_{j=1}^{J} W_j(q-r_j) P(q - r_j,t | q_0, t_0) - \sum_{j = 1}^J W_j(q) P(q,t | q_0, t_0) 
\label{eqn:me}
\end{eqnarray}
where one equation of the form \eqref{eqn:me} is needed for each valid state $q$. Note that the first and second terms on the right-hand side of \eqref{eqn:me} can be interpreted, respectively, as the rates of entering and leaving the state $q$. In order to simplify notation, we will write $P(q,t)$ instead of $P(q,t | q_0, t_0)$ from now on. 

A common application of master equation is to model the dynamics of chemical reactions \cite{Gillespie:1992tq}. We will give a simple example to illustrate that. 

\begin{example} 
Consider the chemical reaction: 
\begin{center}
L + R 
\xrightleftharpoons[\text{k}_{-}]{\text{k}_{+}}
C
\end{center} 
where the chemical species are $L$, $R$ and $C$, and the forward and reverse reaction constants are $k_+$ and $k_-$ respectively. This chemical reaction can be described by a Markov process with state space $q = [ n_L, n_R, n_C]^T$ where $n_L$ is the number of molecules of chemical $L$ etc., and $T$ denotes matrix transpose. 

There are two possible types of jumps (i.e. $J = 2$) in this system. The forward reaction is modelled by the jump $r_1 = [-1, -1 ,1]^T$ where the entries of $r_1$ reflects the fact that one molecule of L and one molecule of $R$ react to form a molecule of $C$. The rate of jump, which according to standard result in chemical kinetics, is $W_1(q) = k_+ q(1) q(2) = k_+ n_L n_R$ where $q(1)$ is the first component of the vector $q$, etc. Similarly, the reverse reaction is modelled by the jump $r_2 = [1, 1, -1]^T$ with $W_2(q) = k_- q(3) = k_- n_C$. The master equation for this chemical reaction is: 
\begin{eqnarray}
\frac{d P(q,t)}{dt} & = & \sum_{j = 1}^2 W_j(q-r_j) P(q - r_j,t) - \sum_{j = 1}^2 W_j(q) P(q,t) 
\label{eqn:cme} 
\end{eqnarray}
Equations of the type \eqref{eqn:cme} is also known as chemical master equations.
\end{example}

\subsection{Results on mean and covariance}
The master equation \eqref{eqn:me} shows the time evolution of the state probability of the Markov process $Q(t)$. However, \eqref{eqn:me} can be difficult to work with because we need one equation for each valid state, hence the number of equations can be very large. Therefore, it is easier if we can determine the mean and covariance of the Markov process $Q(t)$, which is defined as follows: 
\begin{eqnarray}
\langle Q(t) \rangle & = & \sum_{q} q \mbox{Prob}(Q(t) = q) =  \sum_{q} q P(q,t) 
\label{eqn:mean_def}
\\
\Sigma(t) & = & \sum_{q} (q - \langle Q(t) \rangle) (q - \langle Q(t) \rangle)^T P(q,t)
\end{eqnarray} 
where $\langle \bullet \rangle$ will be used in this paper to denote the mean operator. 

It is possible to use \eqref{eqn:me} to derive how the mean and covariance of the state of the Markov process $Q(t)$ evolve over time. 
The following proposition is taken from \cite{Ross:2008up}. 

\begin{proposition}
\label{prop:me}
For the general master equation \eqref{eqn:me}, we have 
\begin{eqnarray}
\frac{d \langle Q(t) \rangle}{dt} & = & \sum_{j=1}^J r_j \langle W_j(Q(t)) \rangle
\end{eqnarray} 

In particular, if $W_j(q)$ is a linear function of $q$. Let $\sum_{j = 1}^J r_j W_j(q) = A q$, then 
\begin{eqnarray}
\frac{d \langle Q(t) \rangle}{dt} & = & A \langle Q(t) \rangle \label{eqn:me_1m} \\
\frac{d \Sigma(t)}{dt} & = & A \Sigma(t) + \Sigma(t) A^T + \sum_{j = 1}^J r_j r_j^T W_j(\langle Q(t) \rangle)
\label{eqn:me_2m} 
\end{eqnarray} 
\end{proposition} 

\section{Modelling molecular communication network using RDMEX} 
\label{sec:rdmex_main} 

\begin{figure}[h]
\begin{center}
\includegraphics[width=10cm]{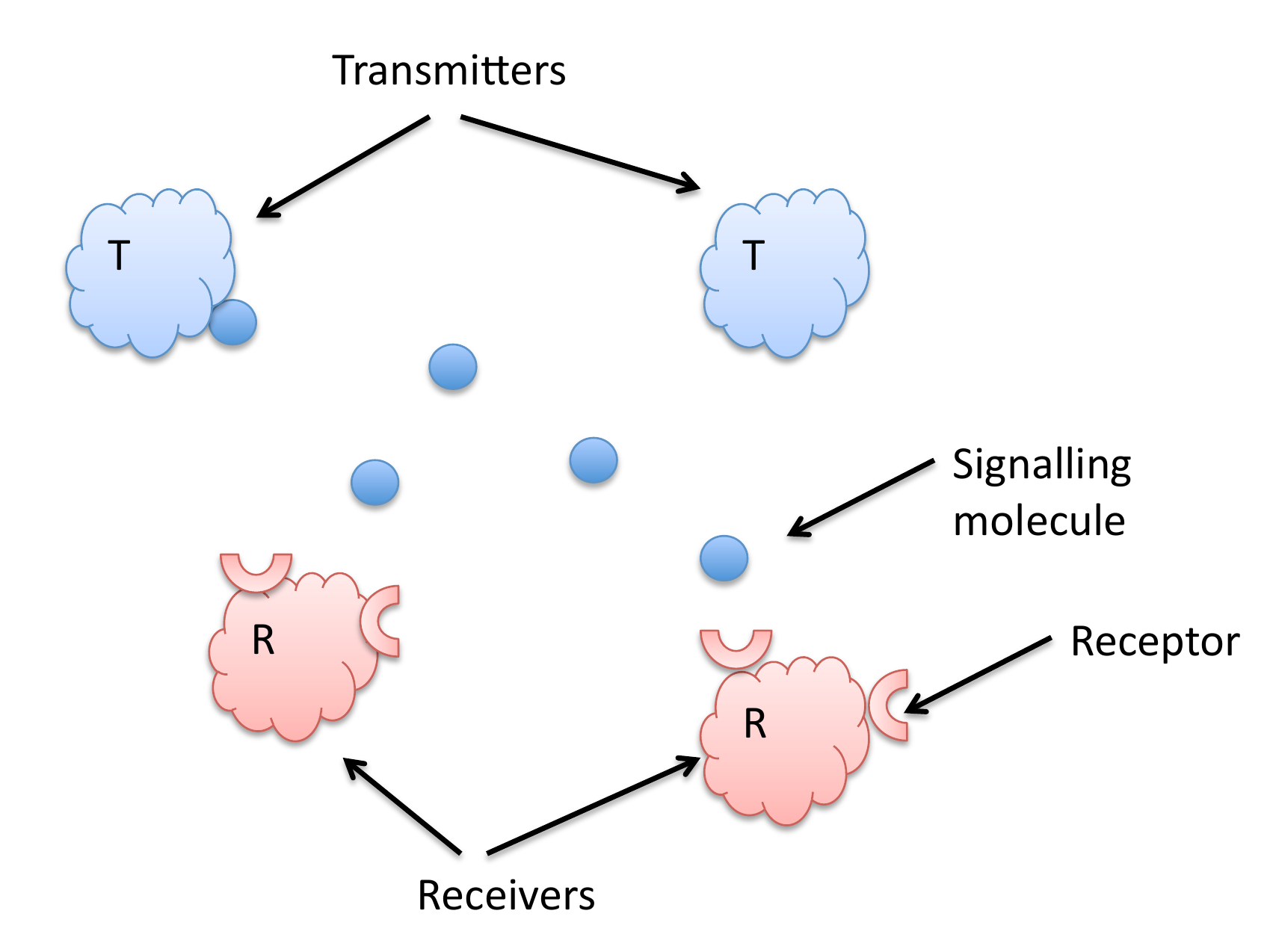}
\caption{An illustration of a molecular communication network with two transmitters and two receivers. The signalling molecules diffuse in a fluidic medium and may bind with the receptors at the receviers.}
\label{fig:mcn}
\end{center}
\end{figure}

We consider a molecular communication network with multiple transmitters and multiple receivers in an isotropic fluidic medium, see Figure \ref{fig:mcn}. The communication takes place by the transmitter emitting one or more signalling molecules over time. Once the molecules leave the transmitter, they diffuse in the medium according to Brownian motion. The receivers are assumed to consist of one or more receptors. When a signalling molecule $L$ reaches a receptor $R$, they may bind together to form a complex $C$. We will consider {\sl the number of complexes at the receiver as the output of the receiver}. For example, a receiver may infer that a bit has been sent by the transmitter when the number of complexes exceeds a threshold. Note that use of this definition of output is also used in chemotaxis models in biophysics \cite{Bialek:2005vq,Endres:2009vx} and molecular communication network models in engineering \cite{Pierobon:2010vg}. 


Based on the above description, we see that a model for molecular communication networks must at least capture the diffusion of signalling molecules and the reactions at the receivers. Both reactions and diffusion can be modelled as Markov processes, so the master equation \eqref{eqn:me} is a natural choice. The remaining issue is how we can model the transmitter. In this paper, we model the transmitter by a time sequence which specifies the number of molecules emitted by the transmitter at a particular time. This approach is fairly general and can be used to model encoding methods that have been considered in the literature, such as molecular coding \cite{Akyildiz:2008vt} and concentration coding \cite{Mahfuz:2011te}. We will consider other modelling approaches, e.g. modelling the internal mechanism of the transmitters, in future work.

In section \ref{sec:rdmex}, we introduce the RDMEX model by way of an example and then we prove some results on mean and covariance of the RDMEX model in section \ref{sec:rdmex_mean}. 

\subsection{The RDMEX model} 
\label{sec:rdmex}

In this section, we will introduce the RDMEX model for a molecular communication network with 2 transmitters, 2 receivers in a 1-dimensional medium. The reason for that is to simplify the presentation. It is fairly simple to generalise the model to multiple transmitters, multiple receivers in a 3-dimensional medium, which we will discuss at the end of the section. 

Another simplification is that we will assume, at the receiver, the rate at which the complexes are formed is a linear function of the number of local signalling molecules and is independent of the number of receptors. (This will be made precise below.) This simplification allows us to produce closed form expressions in the continuum limit. Note that, it is straightforward to model non-linear reaction rates or incorporate more complex receivers in our model.

The following is a list of model assumptions, parameters and notation.

\begin{figure}[h] 
\begin{center}
\includegraphics[width=10cm]{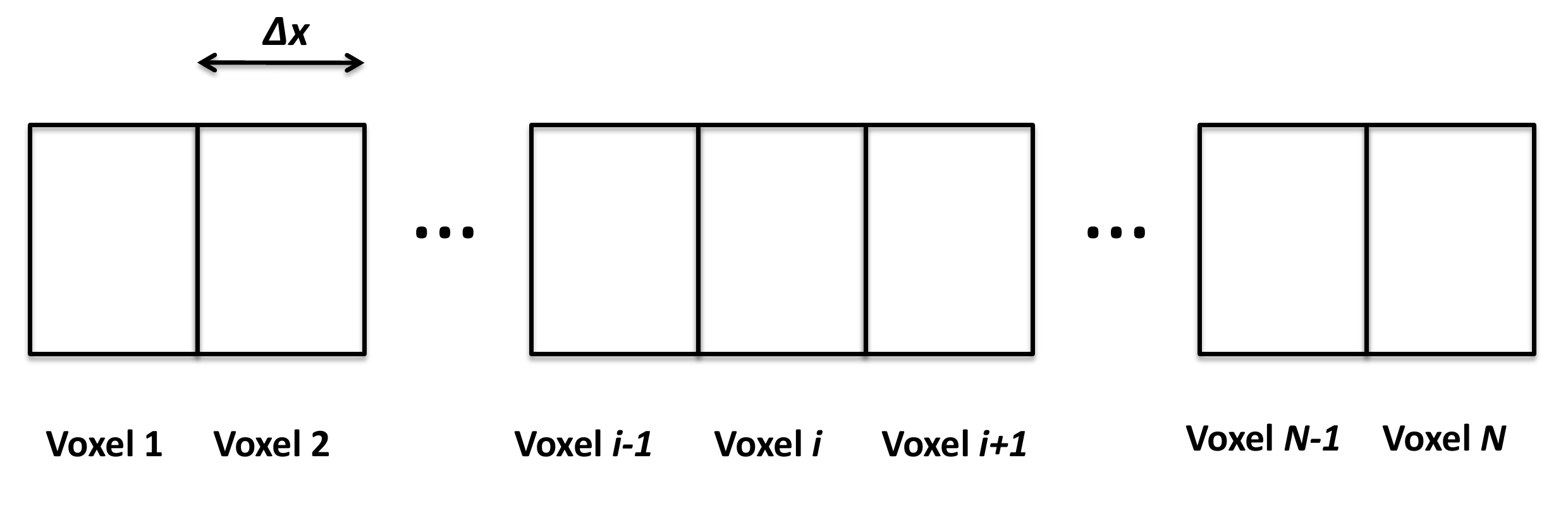}
\caption{The voxels in the 1-dimensional RDMEX model.}
\label{fig:voxel}
\end{center}
\end{figure}

\begin{enumerate}
\item We assume that both transmitters use one and the same type of signalling molecule $L$. 
\item For transmitter 1, we assume that it emits $k_{1,1}$ signalling molecules of $L$ at time $t_{1,1}$, $k_{1,2}$ molecules of $L$ at time $t_{1,2}$, ..., $k_{1,b}$ molecules of $L$ at time $t_{1,b}$, where $k_{1,b}$ $(b = 1,2,3,...)$ are positive integers and $t_{1,b}  \in \mathbb{R}$. Similarly, transmitter 2 emits $k_{2,b}$ signalling molecules of $L$ at time $t_{2,b}$ where $b = 1,2,3,...$. The number of molecules $k_{a,b}$ emitted at time $t_{a,b}$ is assumed to be independent of the state of the system at or before $t_{a,b}$. 
\item The medium is assumed to be a 1-dimensional space of length $X$. The medium is partitioned into $N$ equal width voxels of width $\Delta x$ such that $N \; \Delta x = X$. We index the voxels by using $1,...,N$. See Figure \ref{fig:voxel} for an illustration. 
\item The medium is assumed to be isotropic. The rate at which a signalling molecule $L$ diffuses from one voxel to a neighbouring voxel is $\ddone$ per molecule per unit time. The rate of diffusion from a voxel to a non-neighbouring voxel is zero. Also, the molecule cannot leave the medium and we assume the boundary is reflective. The parameter $\ddone$ is related to one-dimensional macroscopic diffusion constant $\breve{D}$ by $\ddone = \frac{\breve{D}}{\Delta x^2}.$ 
\item We assume that each transmitter or receiver occupies only one voxel. Transmitters 1 and 2 are located respectively in the voxel indexed by $T_1$ and $T_2$. Similarly, we assume that receivers 1 and 2 are located in the voxels indexed by $R_1$ and $R_2$. The indices $T_1$, $T_2$, $R_1$ and $R_2$ are integers in the interval $[1,N]$ and are assumed to be distinct. (Note that it is simple to modify the model so that a transmitter or a receiver occupies multiple voxels.) 
\item We assume both receivers 1 and 2 use the same type of receptors $R$ and these receptors are fixed in space, i.e. do not diffuse. With the simplification mentioned in the introductory part of this section, we assume that the reaction between the signalling molecule $L$ and the complex $C$ is: 
\begin{center}
$L$  
\xrightleftharpoons[\text{k}_{-}]{\text{k}_{+}}
$C$
\end{center} 
where $k_{+}$ and $k_{-}$ are the macroscopic reaction rate constants. With this assumption, the rate of formation of complexes at receiver 1, at any time, is proportional to the number of signalling molecules $L$ in the $R_1$-th voxel. The situation at receiver 2 is similar. 

\begin{remark}
The reaction kinetics assumed above can be viewed as a linearisation of the second order reaction $L + R$ \xrightleftharpoons[]{} $C$. Similar assumption is also used in \cite{Bialek:2005vq,Endres:2009vx} to model receptor kinetics in chemotaxis. 
\end{remark} 

\item The state vector $q$ consists of $(N+2)$ elements where 
\begin{eqnarray}
q & = & \left[ 
\begin{array}{cccccc}
n_{L,1} & n_{L,2} & .... & n_{L,N} & n_{C,1} & n_{C,2} 
\end{array}
\right]
\label{eqn:sv_rdmex_ex}
\end{eqnarray}
where $n_{L,j}$ represents the number of signalling molecules at the $j$-th voxel and $n_{C,u}$ represents the number of complexes at the $u$-th receiver.
\item We define two indicator vectors $\mathbb{1}_{T_1}, \mathbb{1}_{T_2} \in \mathbb{Z}^{N+2}$. The $T_1$-th element of $\mathbb{1}_{T_1}$ is 1 and are otherwise zero. $\mathbb{1}_{T_2}$ is similarly defined. 
\item In order to write down the diffusion and reaction within the molecular communication network, we define the following state transition vectors $r_j$ and transition rates $W_j$. The total number of possible jumps in this system is $J = 2N+4$ where $2N$ of them model diffusion and the rest models reactions at the receivers. We will state these jumps below in four categories. Note that all $r_j \in \mathbb{Z}^{N+2}$ and only the non-zero elements of $r_j$ are stated, and $q$ is the state vector defined in \eqref{eqn:sv_rdmex_ex}. 
\begin{enumerate}
\item The diffusion of $L$ from voxel $j$ to $j+1$, where $1 \leq j \leq N-1$, is modelled by $r_j$ and $W_j(q)$. Specifically, $r_j(j) = -1$, $r_j(j+1) = 1$ and $W_j(q) = \ddone q(j) = \ddone n_{L,j}$. 

{\sl Explanation:} If a signalling molecule diffuses from voxel $j$ to $j+1$, it means the number of signalling molecules in voxel $j$ is decreased by one (hence $r_j(j) = -1$) and that in voxel $j+1$ is increased by one (hence $r_j(j+1) = 1$). The rate at which this particular type of jumps takes place is proportional to the number of molecules in the $j$-th voxel, which is given by the $j$-th element of the state vector $q$. 

For convenience, we define $r_N$ to be a zero vector and $W_N(q) = 0$.

\item For the diffusion of $L$ from voxel $j$ to $j-1$ where $2 \leq j \leq N$, $r_{N+j}(j) = -1$, $r_{N+j}(j-1) = 1$ and $W_{N+j}(q) = \ddone q(j) = \ddone n_{L,j}$. For convenience, we define $r_{N+1}$ to be a zero vector and $W_{N+1}(q) = 0$.

\item For receiver 1, the vector $r_{2N+1}$ and the rate $W_{2N+1}(q)$ are used to model the forward reaction of the conversion of a signalling molecule $L$ to a complex $C$. Specifically, $r_{2N+1}(R_1) = -1$, $r_{2N+1}(N+1) = 1$ and $W_{2N+1}(q) = \frac{k_+}{\Delta x} q(R_1) = \frac{k_+}{\Delta x} n_{L,R_1}$. 

{\sl Explanation:} In the forward reaction, a signalling molecule is removed in the $R_1$-th voxel, hence $r_{2N+1}(R_1) = -1$ and a complex is formed, hence $r_{2N+1}(N+1) = 1$ because the number of complexes at receiver 1 is the $(N+1)$-th element in the state vector \eqref{eqn:sv_rdmex_ex}. The rate $W_{2N+1}(q)$ is proportional to the number of signalling molecules in the $R_1$-th voxel where the receiver is located.

For the reverse reaction, $r_{2N+2}(N+1) = -1$, $r_{2N+2}(R_1) = 1$ and $W_{2N+2}(q) = k_- q(N+1) = k_- n_{C,1}$.
\item For receiver 2, the forward reaction: $r_{2N+3}(R_2) = -1$, $r_{2N+3}(N+2) = 1$ and $W_{2N+3}(q) = \frac{k_+}{\Delta x} q(R_2) = \frac{k_+}{\Delta x} n_{L,R_2}$. For the reverse reaction, $r_{2N+4}(N+2) = -1$, $r_{2N+4}(R_2) = 1$ and $W_{2N+4}(q) = k_- q(N+2) = k_- n_{C,2}$.
\end{enumerate} 
\end{enumerate} 

The RDMEX model for the 2-transmitter 2-receiver molecular communication network is
\begin{eqnarray}
\frac{d P(q,t)}{dt} & =  & \sum_{a = 1}^2 \sum_{b=1}^\infty \{ P(q - k_{a,b} \mathbb{1}_{T_a}) - P(q,t) \}  \delta(t - t_{a,b}) 
\nonumber  \\ 
& & + \sum_{j = 1}^{J} W_{j}(q-r_{j}) P(q - r_{j},t) - \sum_{j = 1}^{J} W_{j}(q) P(q,t) 
\label{eqn:rdmex} 
\end{eqnarray}
where $\delta(t)$ denotes the Dirac delta function. 

Let us, for the time being, assume that the first term on the right-hand side of \eqref{eqn:rdmex} is not there. If this is the case, then \eqref{eqn:rdmex} is of the same form as the master equation \eqref{eqn:me} and the equation models a Markov process. Given this model includes both reaction and diffusion, equation \eqref{eqn:rdmex} without the first term is known in the literature as reaction-diffusion master equation (RDME). 

The novelty of the RDMEX model is the introduction of the first term on the right-hand side of \eqref{eqn:rdmex}. This term can be viewed as a deterministic input because molecules are emitted by the transmitters at pre-determined times. Let us look at this term more closely. At time $t_{a,b}$, the $a$-th transmitter emits $k_{a,b}$ signalling molecules into the $T_a$-th voxel (where the $a$-th transmitter is located). This means that if the system is in the state $q$ just before the time $t_{a,b}$ (denoted as $t_{a,b}^-$), then it will be in state $q + k_{a,b} \mathbb{1}_{T_a}$ just after $t_{a,b}$ (denoted as $t_{a,b}^+$). In addition, we have $P(q,t_{a,b}^-) = P(q + k_{a,b} \mathbb{1}_{T_a},t_{a,b}^+)$, which is modelled by the first term in \eqref{eqn:rdmex}. Note that it is possible to give a stochastic interpretation of $k_{a,b}$, see Remark \ref{re:sto}. 

The deterministic input in \eqref{eqn:rdmex} can be thought of as an external arrival of the system. We will refer to \eqref{eqn:rdmex} as reaction-diffusion master equation with exogenous input, or RDMEX for short. The name is inspired by  time series models such as ARX and ARMAX \cite{Ljung}. 

Note that the RDMEX model is no longer a Markov process due to the deterministic arrivals. However, RDMEX is piecewise Markovian in the sense that, it is Markovian in between two consecutive deterministic arrivals. 

We have given an example of RDMEX for a 2-transmitter 2-receiver model in 1-dimension. The model can be readily generalised to include more transmitters and receivers. In order to generalise the model to 3-dimensional space, we will need to divide the space into 3-dimensional cubic voxels of equal volume. (The use of more complicated geometry is possible, see \cite{Isaacson:2007ui}.) The molecules in a voxel are only allowed to diffuse to any of its neighbouring voxels. This can also be readily be done. Lastly, we remark that it is also possible to use more complex receiver structure or to consider non-isotropic medium. 

\subsection{Mean and covariance of receiver output in the RDMEX model}
\label{sec:rdmex_mean}
We will now generalise the result of Proposition \ref{prop:me} to the case of RDMEX model. 

\begin{proposition}
\label{prop:rdmex}
For the RDMEX model in \eqref{eqn:rdmex}, assuming that $W_j(q)$ is a linear function of $q$. Let $\sum_{j = 1}^J r_j W_j(q) = A q$, then 
\begin{eqnarray}
\frac{d \langle Q(t) \rangle}{dt} & = & A \langle Q(t) \rangle + \sum_{a = 1}^2 \sum_{b=1}^\infty k_{a,b} \mathbb{1}_{T_a}  \delta(t - t_{a,b}) 
\label{eqn:rdmex_1m}  \\ 
\frac{d \Sigma(t)}{dt} & = & A \Sigma(t) + \Sigma(t) A^T + \sum_{j = 1}^J r_j r_j^T W_j(\langle Q(t) \rangle)
\label{eqn:rdmex_2m}
\end{eqnarray} 
\end{proposition} 

\noindent{\bf Proof:} 
For the time evolution on the mean $\langle Q(t) \rangle$, we can start with the derivative of \eqref{eqn:mean_def}: $\frac{d \langle Q(t) \rangle}{dt} = \sum_q q \frac{d P(q,t)}{dt}$ and then use \eqref{eqn:rdmex} for $\frac{d P(q,t)}{dt}$. This is fairly straightforward and uses exactly the same argument as the proof in \cite{Ross:2008up}. 

Alternatively, one can argue the correctness of \eqref{eqn:rdmex_1m} as follows. Given that the difference between \eqref{eqn:me} and \eqref{eqn:rdmex} is the deterministic arrivals modelled by impulses, this means that between two consecutive deterministic arrivals, the evolution of the state in RDMEX can be described by a standard master equation. Hence, \eqref{eqn:me_1m} holds between two consecutive deterministic arrivals. It can be readily shown that the effect of $k_{a,b}$ molecules arriving at time $t_{a,b}$ is to add $k_{a,b} \mathbb{1}_{T_a}$ to the state vector. Hence the form of \eqref{eqn:rdmex_1m}. 

For the evolution of covariance, one can follow the derivation in \cite{Ross:2008up} provided that the impulses are handled correctly because the multiplication of Dirac deltas (or distributions) is not well defined. However, one can argue the correctness of \eqref{eqn:rdmex_2m} using the same argument in the last paragraph. We know that between two consecutive deterministic arrivals, \eqref{eqn:me_2m} holds for the RDMEX model. It remains to show that the covariance matrix just before a deterministic arrival is equal to that just after the deterministic arrival. 

Let $Q(t_{a,b}^-)$ and $\langle Q(t_{a,b}^-) \rangle$ be the state and mean state just before the deterministic arrival at time $t_{a,b}$. At time $t_{a,b}^+$, just after $t_{a,b}$, the state of the system will become $Q(t_{a,b}^-) + k_{a,b} \mathbb{1}_{T_a}$. Also, the mean state at $t_{a,b}^+$ is:
\begin{eqnarray}
\langle Q(t_{a,b}^+) \rangle & = & \sum_q q P(q, t_{a,b}^+) = \sum_q q P(q - k_{a,b} \mathbb{1}_{T_a} , t_{a,b}^-)  \nonumber \\
& = &
\sum_q (q + k_{a,b} \mathbb{1}_{T_a}) P(q, t_{a,b}^-) = \langle Q(t_{a,b}^-) \rangle + k_{a,b} \mathbb{1}_{T_a}
\end{eqnarray}
Note that we have used the fact that $P(q, t_{a,b}^+) = P(q - k_{a,b} \mathbb{1}_{T_a} , t_{a,b}^-)$ in the above derivation. The overall result is that, at time $t_{a,b}$, the state $Q(t)$ and mean state $\langle Q(t) \rangle$ are incremented by the same vector $k_{a,b} \mathbb{1}_{T_a}$. 

Given that, at any deterministic arrival, both the state and mean state change by exactly the same amount, therefore, deterministic arrivals do not cause discontinuity in covariance. Hence \eqref{eqn:rdmex_2m}. \hfill $\Box$

For the rest of the paper, we will focus on studying the properties of equation \eqref{eqn:rdmex_1m}, though we will present a numerical example in Section \ref{sec:num} to demonstrate the accuracy of \eqref{eqn:rdmex_2m}. A detail study on \eqref{eqn:rdmex_2m} is also important and will be done in a future paper. 

\begin{remark}
\label{re:sto} 
We will now briefly discuss a generalisation of the RDMEX model. Instead of assuming a deterministic emission of exactly $k_{a,b}$ signalling molecules by the $a$-th transmitter at time $t_{a,b}$, one may assume that the number of molecules emitted is a random variable $K_{a,b}$ with mean $\langle K_{a,b} \rangle$ and covariance $\mbox{cov}(K_{a,b})$. Provided that the random variable $K_{a,b}$ is independent of the state $q$ at time $t_{a,b}$ or earlier, similar results to Proposition \ref{prop:rdmex} can be derived. For equation \eqref{eqn:rdmex_1m}, we need to replace $k_{a,b}$ by $\langle K_{a,b} \rangle$, and we need to add $\mbox{cov}(K_{a,b})$ to the right-hand side of \eqref{eqn:rdmex_2m}. This generalisation says that one can interpret $k_{a,b}$ in \eqref{eqn:rdmex_1m} as the mean number of molecules emitted at time $t_{a,b}$ by the $a$-th transmitter. With this stochastic interpretation of $k_{a,b}$, one can consider the signalling molecules are generated by an irreversible chemical reaction. 
\end{remark} 

\section{Continuum limit of RDMEX} 
\label{sec:cont} 

In section \ref{sec:rdmex}, we present an example of the RDMEX model for a 2-transmitter, 2-receiver molecular communication network in an isotropic 1-dimensional medium. We also show that if the reaction kinetics at the receiver is linear, then the mean number of molecules in the network evolves according to the ODE \eqref{eqn:rdmex_1m}. In section \ref{sec:gen}, we will determine the continuum limit of \eqref{eqn:rdmex_1m} as $\Delta x \rightarrow 0$. In order to simplify the presentation, we have so far limited our study to 1-dimensional but given most molecular communication networks are 3-dimensional, we will generalise the continuum limit to 3-dimensional case as well. 

The continuum limit of the RDMEX is in fact a RDPDE. A nice property of the resulting RDPDE is that a closed form solution is available. This closed form solution shows that the output signal of a receiver can be affected by the presence of other receivers in the network.  This will be discussed in section \ref{sec:rdpde}. 

\subsection{Continuum limit and generalisation to 3-dimensional space}
\label{sec:gen} 
In this section, we will study the continuum limit of equation \eqref{eqn:rdmex_1m} and show that in the limit, when the interval $\Delta x$ goes to zero, \eqref{eqn:rdmex_1m} converges to a RDPDE and a number of chemical kinetics ODEs. In order that we will be able to solve the RDPDE analytically later on, we will assume from now onwards that the 1-dimensional medium is infinite, which means that the molecules in each voxel can diffuse to either of its neighbouring voxel and the state vector $q$ is 
\begin{eqnarray}
q & = & \left[ 
\begin{array}{ccccccccc}
 .... & n_{L,-2} & n_{L,-1} & n_{L,0} & n_{L,1} & n_{L,2} & ....  & n_{C,1} & n_{C,2} 
\end{array}
\right]
\end{eqnarray}
where, as before, $n_{L,j}$ is the number of $L$ in the $j$-th voxel where $j \in \mathbb{Z}$, and $n_{C,u}$ is the number of complexes formed at the $u$-th receiver. Given this state vector, we can write equation \eqref{eqn:rdmex_1m} as
\begin{eqnarray}
\frac{d \langle n_{L,j}(t) \rangle}{dt} & = & \ddone (\langle n_{L,j-1}(t) \rangle - 2  \langle n_{L,j}(t) \rangle + \langle n_{L,j+1}(t) \rangle) + \sum_{a = 1}^2 \sum_{b = 1}^\infty \delta_K(j - T_a)  k_{a,b} \delta(t - t_{a,b}) \nonumber \\ 
& & - \sum_{u = 1}^2 \delta_K(j - R_u) ( \frac{k_+}{\Delta x}  \langle n_{L,R_u}(t) \rangle - k_-  \langle n_{C,u}(t) \rangle )  \; \; \forall j \in \mathbb{Z} \label{eqn:m1d_int} \\
\frac{d \langle n_{C,u}(t) \rangle}{dt} & = & \frac{k_+}{\Delta x}  \langle n_{L,R_u}(t) \rangle - k_-  \langle n_{C,u}(t) \rangle \mbox{ for } u = 1,2 
 \label{eqn:m1d_c}
\end{eqnarray} 
where $\delta_K(j)$ is the Kronecker delta\footnote{Note: We use both Konecker delta and Dirac delta in this paper. They are denoted, respectively, as $\delta_K(\bullet)$ and $\delta(\bullet)$.} . 

Suppose the centre of the voxel $j$ is at position $x_j$, we replace $\langle n_{L,j}(t) \rangle$ by $\ell(x_j,t) \Delta x$ where $\ell(x_j,t)$ is the mean concentration in voxel $j$ at time $t$. By dividing both sides of \eqref{eqn:m1d_int} by $\Delta x$, taking the limit $\Delta x \rightarrow 0$ and noting that $\ddone (\Delta x)^2 = \breve{D}$, we have 

\begin{eqnarray}
\frac{\partial \ell }{\partial t} & = & \breve{D} \frac{\partial^2 \ell}{\partial x^2}  + \sum_{a = 1}^2 \delta(x - x_{T,a}) 
\underbrace{\sum_{b = 1}^\infty   k_{a,b} \delta(t - t_{a,b})}_{= k_a(t)}  
- \sum_{u = 1}^2 \delta(x - x_{R,u}) \frac{d \langle n_{C,u}(t) \rangle}{dt} \label{eqn:rdpde_1d} \\
\frac{d \langle n_{C,u}(t) \rangle}{dt} & = &  k_+  \ell(x_{R,u},t) - k_-  \langle n_{C,u}(t) \rangle \mbox{ for } u = 1,2 
\label{eqn:rd_1d}
\end{eqnarray} 
where $x_{T,a}$ (resp. $x_{R,u}$) is the centre of the voxel $T_a$ ($R_u$) where the $a$-th transmitter ($u$-th receiver) is located. Note that we have also used the following conversion between the Kronecker and Dirac deltas: $\lim_{\Delta x \rightarrow 0} \frac{\delta_K(j)}{\Delta x} = \delta(x)$. 

This shows that in the continuum, the RDMEX converges to a RDPDE \eqref{eqn:rdpde_1d} and a number of ODEs describing the kinetics at the receivers \eqref{eqn:rd_1d}. The RDPDE \eqref{eqn:rdpde_1d} has a simple interpretation. The second term in \eqref{eqn:rdpde_1d} says that the transmitter at $T_a$ adds signalling molecules to the system according to time sequence $k_a(t)$, while the third terms says the signalling molecules are absorbed from the system if they form complexes at the receivers. Given that we assume that one signalling molecule reacts to form one complex, therefore the rate of absorption of signalling molecules is equal to the rate of complex formation, which is given by \eqref{eqn:rd_1d}. 

Note that it is well known in literature, see \cite{Gardiner,Isaacson:2007ui}, that a RDME with linear reaction rates converges to a RDPDE in continuum. In the above, we show analogues result holds for the RDMEX model. 

\subsubsection{Generalisation to 3-dimensional space}
In order to simplify the presentation, we have so far limited to the 1-dimensional case. Given that most molecular communication networks are 3-dimensional, we state that in a 3-dimensional infinite medium the RDMEX will converge to the following RDPDE in the continuum. Here $v$ denotes a point in the 3-dimensional space (i.e.~$v$ is a 3-dimensional vector) and $\ell(v,t)$ is the mean concentration of the signalling molecule at the location $v$ at time $t$. 

\begin{eqnarray}
\frac{\partial \ell }{\partial t} & = & D \nabla^2 \ell  + \sum_{a = 1}^2 \delta(v - v_{T,a}) k_a(t) 
- \sum_{u = 1}^2 \delta(v - v_{R,u}) \frac{d c_u(t)}{dt} \label{eqn:rdpde_3d} \\
\frac{d c_u(t)}{dt} & = &  k_+  \ell(x_{R,u},t) - k_-  c_u(t) \mbox{ for } u = 1,2 
\label{eqn:rd_3d}
\end{eqnarray} 
where $\nabla^2$ is the Laplacian in 3-dimensional space, and $v_{T,a}$ (resp. $v_{R,u}$) is a 3-dimensional vector specifying the location of $a$-th transmitter ($u$-th receiver). Note that we have also introduced a new notation $c_u(t)$ to denote the mean number of complexes $\langle n_{C,u}(t) \rangle$ at receiver $u$; this is to simplify the notation later. 

The derivation assumes that the molecule in a voxel diffuses to a neighbouring voxel at a rate of $\dd$ per molecule per unit time and each voxel is a cube of size $\Delta^3$. The parameter $\dd$ is related to the 3-dimensional macroscopic diffusion constant $D$ by $\dd = \frac{D}{\Delta^2}$. Also, the rate of formation of complexes at the receiver voxel is given by $\frac{k_+}{\Delta^3}$ times the number of signalling molecules $L$ in the receiver voxel. 
Given that the derivation for the 3-dimensional is essentially the same as the 1-dimensional case, we do not present it here. 

\subsection{Solution to RDPDE}
\label{sec:rdpde}
In this section, we present a solution to the RDPDE \eqref{eqn:rdpde_3d} and \eqref{eqn:rd_3d}, which is the continuum limit of the 3-dimensional RDMEX model. The key result is a closed-form expression of the multivariate transfer function from the transmitter signals $k_1(t)$ and $k_2(t)$ (which models the number of molecules injected by the transmitters into the medium at time $t$ and can be viewed as the inputs to the system) to the mean number of complexes formed at the receivers $c_1(t)$ and $c_2(t)$ (which can be viewed as the outputs). We will divide this section into two parts. We will first derive the transfer function and then provide an interpretation of the transfer function. 


\subsubsection{Derivation of transfer function} 
The aim of this part is to derive a multivariate transfer function from $k_1(t)$ and $k_2(t)$ to $c_1(t)$ and $c_2(t)$ using \eqref{eqn:rdpde_3d} and \eqref{eqn:rd_3d}. 

We first define a few notation. Let $\iota = \sqrt{-1}$, and $C_u(\omega)$, $K_a(\omega)$ and $\tilde{L}(v,\omega)$ be the temporal Fourier transform of, respectively, $c_u(t)$, $k_a(t)$ and $\ell(v,t)$ where $\omega$ is the transform variable. It is shown in Appendix \ref{app:rdpde} that 
\begin{eqnarray}
\tilde{L}(v,\omega) & = & \sum_{a=1}^2 \phi(v - v_{T,a},\omega) K_a(\omega) - \sum_{u=1}^2 \phi(v - v_{R,u},\omega) \iota \omega C_u(\omega)
\label{eqn:Ltilde}
\end{eqnarray} 
where 
\begin{eqnarray}
\phi(v,\omega) & = & \frac{1}{4 \pi D \| v \|} \exp(-\sqrt{\frac{\iota \omega}{D}} \| v \|)
\label{eqn:phi} 
\end{eqnarray} 
is the temporal Fourier transform of the 3-dimensional diffusion kernel $\frac{1}{(4 \pi D t)^{\frac{3}{2}}} \exp(- \frac{\| v \|^2}{4 D t})$. Given \eqref{eqn:Ltilde} holds for any location $v$, we can use it to determine the concentration of the signalling molecules at the two receivers. By substituting $v = v_{R,1}$ and then $v = v_{R,2}$ in \eqref{eqn:Ltilde}, we have:
\begin{eqnarray}
\tilde{L}(v_{R,1},\omega) & = & \phi_{11}(\omega) K_1(\omega) + \phi_{12}(\omega) K_2(\omega) - \phi_0(\omega) \iota \omega C_1(\omega) - \phi_{\Delta R}(\omega) \iota \omega C_2(\omega) 
\label{eqn:cr1_omega} \\
\tilde{L}(v_{R,2},\omega) & = & \phi_{21}(\omega) K_1(\omega) + \phi_{22}(\omega) K_2(\omega) - \phi_{\Delta R}(\omega) \iota \omega C_1(\omega) - \phi_{0}(\omega) \iota \omega C_2(\omega)
\label{eqn:cr2_omega} 
\end{eqnarray} 
where 
\begin{eqnarray}
\phi_{ua}(\omega) & = & \phi(v_{R,u} - v_{T,a},\omega) \mbox{ for } u,a = 1,2 \\
\phi_{\Delta R}(\omega) & = & \phi(v_{R,1} - v_{R,2},\omega) \\
\phi_{0}(\omega) & = & \phi(\mathbf{0},\omega) \mbox{ where } \mathbf{0} = \mbox{ zero vector}
\label{eqn:phi0} 
\end{eqnarray}

One can interpret $\phi_{ua}(\omega)$ as the transfer function which models the dynamics of the diffusion of molecules from the $a$-th transmitter located at $v_{T,a}$ (where the molecules are injected into the medium) to the $u$-th receiver located at $v_{R,u}$. 

When signalling molecules are absorbed to form complexes, it creates a concentration gradient. The diffusion dynamics between the two receivers are modelled by $\phi_{\Delta R}(\omega)$. Given that we assume that the locations of the transmitters and receivers are distinct, both $\phi_{ua}(\omega)$ and $\phi_{\Delta R}(\omega)$ are well defined. 

The transfer function $\phi_{0}(\omega)$ models the local impact of absorption of signalling molecule at each receiver. This transfer function is unfortunately not well defined as can be seen from substituting $v = \mathbf{0}$ in the definition of $\phi(v,\omega)$ in \eqref{eqn:phi}. The transfer function $\phi_{0}(\omega)$ also appears in the modelling of receptor noise in chemotaxis in the biophysics literature \cite{Bialek:2005vq,Endres:2009vx}. In fact \cite[Equation (18)]{Bialek:2005vq} and \cite[Equation (6)]{Endres:2009vx} are special cases of \eqref{eqn:rdpde_3d} where the input terms $k_a(t)$ are absent. Both \cite{Bialek:2005vq,Endres:2009vx} deal with the indefiniteness of $\phi_{0}(\omega)$ by cutting off an integral to evaluate $\phi_{0}(\omega)$ at a finite frequency. However, this requires us to make an assumption on the size of the receptor molecule. Instead, in section \ref{sec:zt}, we derive a new method to approximate $\phi_{0}(\omega)$, and we will show using numerical examples in section \ref{sec:num} that our approximation gives accurate results. For the rest of this section, we will continue to use equations \eqref{eqn:cr1_omega} and \eqref{eqn:cr2_omega} with the understanding that $\phi_{0}(\omega)$ is not well defined but can be well approximated. 

Both equations \eqref{eqn:cr1_omega} and \eqref{eqn:cr2_omega} are obtained from \eqref{eqn:rdpde_3d}. We still need to work on \eqref{eqn:rd_3d}. By taking the Fourier transform of \eqref{eqn:rd_3d}, we have
\begin{eqnarray}
\rho_1(\omega) & = & \frac{C_1(\omega)}{\tilde{L}(v_{R,1},\omega)}  = \frac{k_+}{\iota \omega + k_-} \label{eqn:ft_cr1} \\
\rho_2(\omega) & = & \frac{C_2(\omega)}{\tilde{L}(v_{R,2},\omega)}  = \frac{k_+}{\iota \omega + k_-} \label{eqn:ft_cr2}
\end{eqnarray}
where $\rho_1(\omega)$ and $\rho_2(\omega)$ are transfer functions that model the reaction kinetics at the receivers. Given that we have assumed that the binding and unbinding rate constants at both receivers are identical, it is not surprising that $\rho_1(\omega)$ and $\rho_2(\omega)$ are the same here. It is straightforward to generalise to the case where the receivers have different reaction kinetics. 

By using equations \eqref{eqn:cr1_omega}, \eqref{eqn:cr2_omega}, \eqref{eqn:ft_cr1} and \eqref{eqn:ft_cr2}, we can eliminate $\tilde{L}(v_{R,1},\omega)$ and $\tilde{L}(v_{R,2},\omega)$ to obtain the transfer function from the inputs $k_1(t)$ and $k_2(t)$ to the outputs $c_1(t)$ and $c_2(t)$:
\begin{eqnarray}
\left[ \begin{array}{c} C_1 \\ C_2 \end{array} \right] 
& = & 
\left( I + \iota \omega 
\underbrace{\left[ \begin{array}{cc} \rho_1 & 0 \\ 0 & \rho_2 \end{array} \right]}_{\cal R} 
\underbrace{\left[ \begin{array}{cc} \phi_0 & \phi_{\Delta r} \\ \phi_{\Delta r}  & \phi_{0} \end{array} \right]}_{\Phi_0}
\right)^{-1}
\left[ \begin{array}{cc} \rho_1 & 0 \\ 0 & \rho_2 \end{array} \right] 
\underbrace{\left[ \begin{array}{cc} \phi_{11} & \phi_{12} \\ \phi_{21}  & \phi_{22} \end{array} \right]}_{\Phi}
\left[ \begin{array}{c} K_1 \\ K_2 \end{array} \right] 
\label{eqn:tf} 
\end{eqnarray}
where, for conciseness, we have dropped the dependence on transform variable $\omega$. Equation \eqref{eqn:tf} is the key result of this section. It is the solution to the RDPDE \eqref{eqn:rdpde_3d} and \eqref{eqn:rd_3d}, which are in turn the continuum limit of the mean concentration in the stochastic RDMEX model. We will provide some physical interpretation of \eqref{eqn:tf} in a moment. Before that, we want to point out that \eqref{eqn:tf} can be used to compute $c_1(t)$ and $c_2(t)$ given $k_1(t)$, $k_2(t)$ and the system parameters by numerical Laplace transform. This will be done in Section \ref{sec:num}. 

Note that it is numerically more efficient to solve for the receiver outputs using \eqref{eqn:tf} rather than \eqref{eqn:rdmex_1m}. This is because one also needs to solve for the number of signalling molecules in the voxels in \eqref{eqn:rdmex_1m} but this is not needed when \eqref{eqn:tf} is used. Since the number of voxels is much larger than the number of transmitters and receivers, numerical solution via \eqref{eqn:tf} is more efficient. 

\subsubsection{Interpretation of transfer function \eqref{eqn:tf}} 
\label{sec:tf_int}
Equation \eqref{eqn:tf} may not look easy to interpret in the first instance, so we will first specialise it to the case of 1-transmitter and 1-receiver. In this case, we have
\begin{eqnarray}
C_1 & = & \frac{\rho_1 \phi_{11} }{1 + \iota \omega \phi_0 \rho_1}  K_1 
\label{eqn:tf11} 
\end{eqnarray}

\begin{figure}
\begin{center}
\includegraphics[width=10cm]{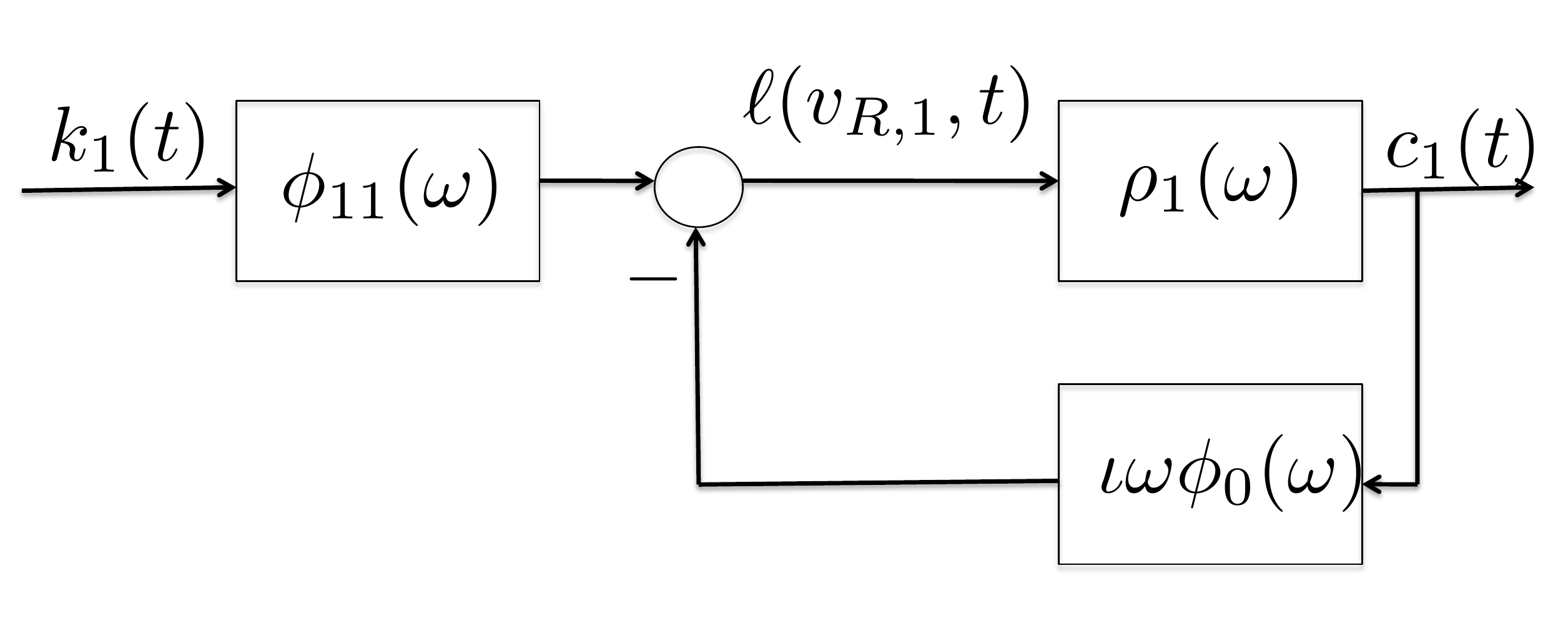}
\caption{Block diagram showing the transfer function from the input $k_1(t)$ to the output $c_1(t)$ of a molecular communication network with one transmitter and one receiver.}
\label{fig:block11}
\end{center}
\end{figure}

One can readily show that the input-output transfer function in \eqref{eqn:tf11} corresponds to the block diagram representation of figure \ref{fig:block11}. The negative feedback loop occurs because the net number of signalling molecules at the receiver $\ell(v_{R,1},t)$ is given by the difference between those that arrive via diffusion (modelled by the feedforward block of $\phi_{11})$ minus those reacted to form the complexes (modelled by the feedback block of $\iota \omega \phi_0$). The forward loop consists of $\phi_{11}$ which models the diffusion dynamics of signalling molecule from the location of the transmitter to that of the receiver, and $\rho_1$ which models the conversion of the signalling molecules to complexes. 

We will now take a closer look at the denominator of \eqref{eqn:tf11} with the aim to determine the strength of the feedback term $\iota \omega \phi_0 \rho_1$. It has been shown in \cite{Bialek:2005vq,Endres:2009vx} that $\phi_0 \propto \frac{1}{2 \pi D}$ at low frequency. Consider the case where $D \gg k_+$ or $\frac{k_+}{D} \approx 0$, which corresponds to the situation where the chemical kinetics is not diffusion-limited. We have $\iota \omega \phi_0 \rho_1 \approx 0$ base on the expression of $\rho_1$ in \eqref{eqn:ft_cr1} and consequently $C_1 \approx \rho_1 \phi_{11} K_1$. Since the chemical kinetics is not diffusion-limited, the chemical kinetics and diffusion are basically "decoupled", so the transfer function from $K_1$ to $C_1$ is the multiplication of the transfer function $\phi_{11}$ from $K_1$ to $\tilde{L}(v_{R_1},\omega)$  (which models diffusion) and the transfer function $\rho_1$ from $\tilde{L}(v_{R_1},\omega)$ to $C_1$ (which models reaction kinetics). We will show in section \ref{sec:num} using numerical examples to show that the transfer function $\frac{\rho_1 \phi_{11} }{1 + \iota \omega \phi_0 \rho_1}  \approx \rho_1 \phi_{11}$ holds when $D \gg k_+$. 


Given the interpretation of the 1-transmitter 1-receiver case as a system with feedback in figure \ref{fig:block11}, it can be shown that equation \eqref{eqn:tf} corresponds to multivariate feedback system with 2 inputs and 2 outputs. The block structure of the multivariate feedback system is the same as that in figure \ref{fig:block11} but we need to replace the single-input single-output transfer functions $\phi_{11}$, $\rho_1$ and $\phi_0$ by their multivariate counterparts of $\Phi_1$, $\cal{R}$ and $\Phi_0$ in \eqref{eqn:tf}. 

For the 2-transmitter 2-receiver case, we see from \eqref{eqn:tf} that the response at each receiver is affected by both transmitters, as well as by the other receiver. Let us assume that the transmitters and receivers form two communication pairs where transmitters 1 and 2 intend to communicate with, respectively, receivers 1 and 2. We want to determine the unintended signal at the receivers. Let us assume for the time being that the two receivers are sufficiently far apart so that $\phi_{\Delta r}(\omega)$ is negligible compared with $\phi_0(\omega)$. In addition, we assume the two receivers are identical, so $\rho_1(\omega) = \rho_2(\omega)$. In this case, we can simplify \eqref{eqn:tf} to 
\begin{eqnarray}
\left[ \begin{array}{c} C_1 \\ C_2 \end{array} \right] 
& \approx & 
 \frac{\rho_1}{1 + \iota \omega \phi_0 \rho_1} 
\left[ \begin{array}{cc} \phi_{11} & \phi_{12} \\ \phi_{21}  & \phi_{22} \end{array} \right]
\left[ \begin{array}{c} K_1 \\ K_2 \end{array} \right] 
\end{eqnarray}
Comparing with the single-transmitter single-receiver transfer function in \eqref{eqn:tf}, we can see that the unintended signal due to transmitter 2 on receiver 1 is $\frac{\rho_1 \phi_{12} }{1 + \iota \omega \phi_0 \rho_1}  K_2$. One can readily see that the magnitude of this unintended signal can be reduced if transmitter 2 is well separated in space from receiver 1 because $\phi$ is a decreasing function of distance. Similar conclusion can be drawn for transmitter 1 and receiver 2. Note that the above argument requires that the receivers are well separated. If this is not the case, the matrix inverse in \eqref{eqn:tf} will create a complicated mixture of signal at both receivers. 

In general, spatial separation is a strategy to reduce the magnitude of unintended signal that one communication pair has on the others. It is interesting to point out that \eqref{eqn:tf} allows one to explore other methods to reduce the magnitude of the unintended signal. An interesting case to study is if the transmitters emit molecules at different frequencies and the receivers are frequency sensitive. We will not explore this further here and leave this for future work. 

\section{Discrete solution for mean concentration in RDMEX} 
\label{sec:zt}
In the last section, we presented a closed-form solution to the RDPDE \eqref{eqn:rdpde_3d}. A problem that we face is that the frequency response $\phi_0(\omega)$ in \eqref{eqn:phi0} is not well defined. This problem arises because the size of voxel, in the continuum limit, becomes zero. Therefore, a solution to overcome this problem is to consider finite voxel size instead. This means that we need to work with the 3-dimensional analogue of equations \eqref{eqn:m1d_int} and \eqref{eqn:m1d_c}. 

Consider an isotropic 3-dimensional space. We divide the space into identical cubic voxels of volume $\Delta^3$ each. We index the voxel using a 3-dimensional vector $\xi = [i \; j \; k]$ where $i, j, k \in \mathbb{Z}$; note that we will use $\xi$ and $[i \; j \; k]$ interchangeably in this section. We use $\xi_{T,a}$ ($\xi_{R,u}$) to index the voxel that the $a$-th transmitter ($u$-th receiver) is located. We will also use $\ell_{R,u}(t)$ to denote the mean concentration of signalling molecules at the voxel where the $u$-th receiver is located. Let $\tilde{c}_u(t)$ denote the mean number of complexes at the $u$-th receiver for this discrete model and $\tilde{C}_u(\omega)$ be its continuous Fourier transform. (Just to avoid any possible confusion. We discretise only space, not time. So, $t$ remains continuous.) 

Let $\ell_{i,j,k}(t)$ denotes the mean concentration of the signalling molecule in voxel $i,j,k$. The mean concentration in a voxel is given by the mean number of molecules divided by the volume of a voxel which is $\Delta^3$. One can show that the generalisation of \eqref{eqn:m1d_int} and \eqref{eqn:m1d_c} to 3-dimensional space --- with mean concentration of signalling molecules, rather than mean number, per voxel --- is:  
\begin{eqnarray}
\frac{d \ell_{i,j,k}(t) }{dt} & = & \dd  (\ell_{i-1,j,k}(t) - 2 \ell_{i,j,k}(t) + \ell_{i+1,j,k}(t) ) + \nonumber \\ 
& & \dd (\ell_{i,j-1,k}(t)  - 2 \ell_{i,j,k}(t) + \ell_{i,j+1,k}(t) ) + \nonumber \\ 
& & \dd (\ell_{i,j,k-1}(t)  - 2 \ell_{i,j,k}(t) + \ell_{i,j,k+1}(t) ) + \nonumber \\ 
& & \sum_{a = 1}^2 \frac{1}{\Delta^3} \delta_K(\xi - \xi_{T,a}) k_a(t) - \sum_{u = 1}^2 \frac{1}{\Delta^3} \delta_K(\xi - \xi_{R,u})  \frac{d \bar{c}_u(t)}{dt} \label{eqn:mean_3d_1} \\
\frac{d \tilde{c}_u(t)}{dt} & = & {k_+}  \ell_{R,u}(t) - k_-  \tilde{c}_u(t) \mbox{ for } u = 1,2 
\label{eqn:mean_3d_2}
\end{eqnarray} 
where $D =  \frac{\dd}{\Delta^2}$. One can readily show that the continuum limit of \eqref{eqn:mean_3d_1} and \eqref{eqn:mean_3d_2} is \eqref{eqn:rdpde_3d} and \eqref{eqn:rd_3d} 

Let $\tilde{L}_d(\xi,\omega)$ denote the temporal Fourier transform of $\ell_{i,j,k}(t)$. (Recall that $\xi = [i,j,k]$.) It is shown in Appendix \ref{app:z} that 

\begin{eqnarray}
\tilde{L}_d(\xi,\omega) & = & \sum_{a=1}^2 \psi(\xi - \xi_{T,a},\omega) K_a(\omega) - \sum_{u=1}^2 \psi(\xi - \xi_{R,u},\omega) \iota \omega \tilde{C}_u(\omega)
\label{eqn:Ltilde_d}
\end{eqnarray} 
where 
\begin{eqnarray}
\psi(\xi = [i,j,k] ,\omega) & = & \frac{1}{4 \pi^2 \tilde{D} \Delta} \oint_{\cal C} \oint_{\cal C}  
\frac{W_{z\ast}^{|k|+1}}{W_{z\ast}^2-1} W_x^{i-1} W_y^{j-1} dW_x dW_y
\label{eqn:psi} 
\end{eqnarray}
where $W_x$ and $W_y$ are complex variables and the contour ${\cal C}$ is the unit circle on the complex plane; also, $W_{z\ast}$ is the solution of the following quadratic equation in $W_z$ with modulus less than unity: 
\begin{eqnarray}
W_z^2 - (2 + (W_x - W_x^{-1})^2 + (W_y - W_y^{-1})^2 + \iota \omega \frac{\Delta^2}{{D}}) W_z + 1 = 0
\label{eqn:qe} 
\end{eqnarray}

Note that \eqref{eqn:Ltilde_d} is the discrete space analogue of \eqref{eqn:Ltilde}. Thus one can identify $\phi(\mathbf{0},\omega)$ with $\psi(\mathbf{0},\omega)$. In addition, one can show that: 

\begin{eqnarray}
\left[ \begin{array}{c} \bar{C}_1 \\ \bar{C}_2 \end{array} \right] 
& = & 
\left( I + \iota \omega 
\left[ \begin{array}{cc} \rho_1 & 0 \\ 0 & \rho_2 \end{array} \right]
\left[ \begin{array}{cc} \psi_0 & \psi_{\Delta r} \\ \psi_{\Delta r}  & \psi_{0} \end{array} \right]
\right)^{-1}
\left[ \begin{array}{cc} \rho_1 & 0 \\ 0 & \rho_2 \end{array} \right] 
\left[ \begin{array}{cc} \psi_{11} & \psi_{12} \\ \psi_{21}  & \psi_{22} \end{array} \right]
\left[ \begin{array}{c} K_1 \\ K_2 \end{array} \right] 
\label{eqn:tf_z} 
\end{eqnarray}
where for conciseness we have dropped the dependence on $\omega$, and $\psi_{ua}(\omega) = \psi(\xi_{R,u}-\xi_{T,a},\omega)$, $\psi_{\Delta r}(\omega) = \psi(\xi_{R,1} - \xi_{R,2},\omega)$ and $\psi_0(\omega) = \psi(\mathbf{0},\omega)$.  

Numerical integration can be used to compute $\psi(\xi,\omega)$ in \eqref{eqn:psi}. This will be used in Section \ref{sec:num}. 

\section{Numerical examples}
\label{sec:num} 
\subsection{Overview} 
In this section, we will give a number of numerical examples to show that equations \eqref{eqn:tf} in section \ref{sec:rdpde} and \eqref{eqn:tf_z} in section \ref{sec:zt} can be used to accurately predict the mean output of the receivers in the stochastic model RDMEX. We will also use these numerical examples to illustrate the issues of using molecular signalling for communication. We will present three sets of results: single-transmitter single-receiver in section \ref{sec:1t1r}, single-transmitter two-receiver in section \ref{sec:1t2r} and two-transmitter two-receiver in \ref{sec:2t2r}. 

In order to verify the accuracy of \eqref{eqn:tf} and \eqref{eqn:tf_z}, we will use simulation to compute the mean receiver output. One method is to simulate RDMEX many times and compute the mean. Alternatively, one can use the fact that if the number of molecules is large, then the behaviour of one simulation run is fairly close to the mean. We will mainly use the latter method in this paper. We simulate the RDMEX model using the $\tau$-leaping method \cite{Gillespie:2007uq}. The $\tau$-leaping algorithm uses a constant time step to advance the simulation and is a faster alternative to the Gillespie's algorithm \cite{Gillespie:2007uq}. We will refer to the simulation result as RDMEX-$\tau$. 

The fluidic medium is assume to have a $D$ of 0.05. (Since the parameters in the diffusion-reaction system can be scaled to some dimensionless quantities \cite[Section 8.2]{Crank}, we do not specify the units for the parameters here.) The locations of the transmitters and receivers, as well as the reaction rate constants, vary between experiments and will be specified later. Different transmitter signals will be used to demonstrate the accuracy of the RDMEX model. Our goal is to compare the output $c_1(t)$ and $c_2(t)$ from RDMEX-$\tau$ with that from the following analytical models: 

\begin{enumerate}
  \item Equation \eqref{eqn:tf_z}. We will refer to this as RDMEX-M.
  \item Equation \eqref{eqn:tf} of the continuum model with $\phi(\mathbf{0},\omega)$ replaced by $\psi(\mathbf{0},\omega)$. We will refer to this as RDPDE. 
  \item Equation \eqref{eqn:tf} of the continuum model with $\phi(\mathbf{0},\omega)$ approximated by $\frac{1}{2 \pi D \Delta}$. This approximation is inspired by the one used in \cite{Bialek:2005vq,Endres:2009vx} where we have replaced the size of receptors by the voxel dimension parameter $\Delta$. We will refer to this as RDPDE-X. 
  \item For the single-transmitter and single-receiver case, we consider the ``decoupled" model $C_1 = \rho_1 \phi_{11} K_1$. This will be referred to as DE. 
\end{enumerate}

For all these analytical models, we first determine the Laplace transforms $C_1$ and $C_2$ by using the Laplace transforms $K_1$ and $K_2$, and the transfer function. We then invert the Laplace transform numerically using the matlab function \verb|invlap.m| \cite{inverseL}. 

\subsection{Single-transmitter single-receiver case}
\label{sec:1t1r} 
The system consists of a transmitter at the voxel $[0,0,0]$ and a receiver at voxel $[3,0,0]$. 

\subsubsection{Model accuracy and the effect of $k_+$} 
\label{sec:1t1r_kp}
In this set of experiments, the transmitter emits 10 molecules every $10^{-4}$ time units for a duration of 0.2 time units and then stops emitting for 0.3 time units. The signal $k_1(t)$ is obtained by concatenating this emission pattern 3 times. The value of $k_-$ is $0.05$. We determine the output signal $c_1(t)$ for $t \in [0,2]$. 

\begin{figure}
\begin{center}
\includegraphics[width=10cm]{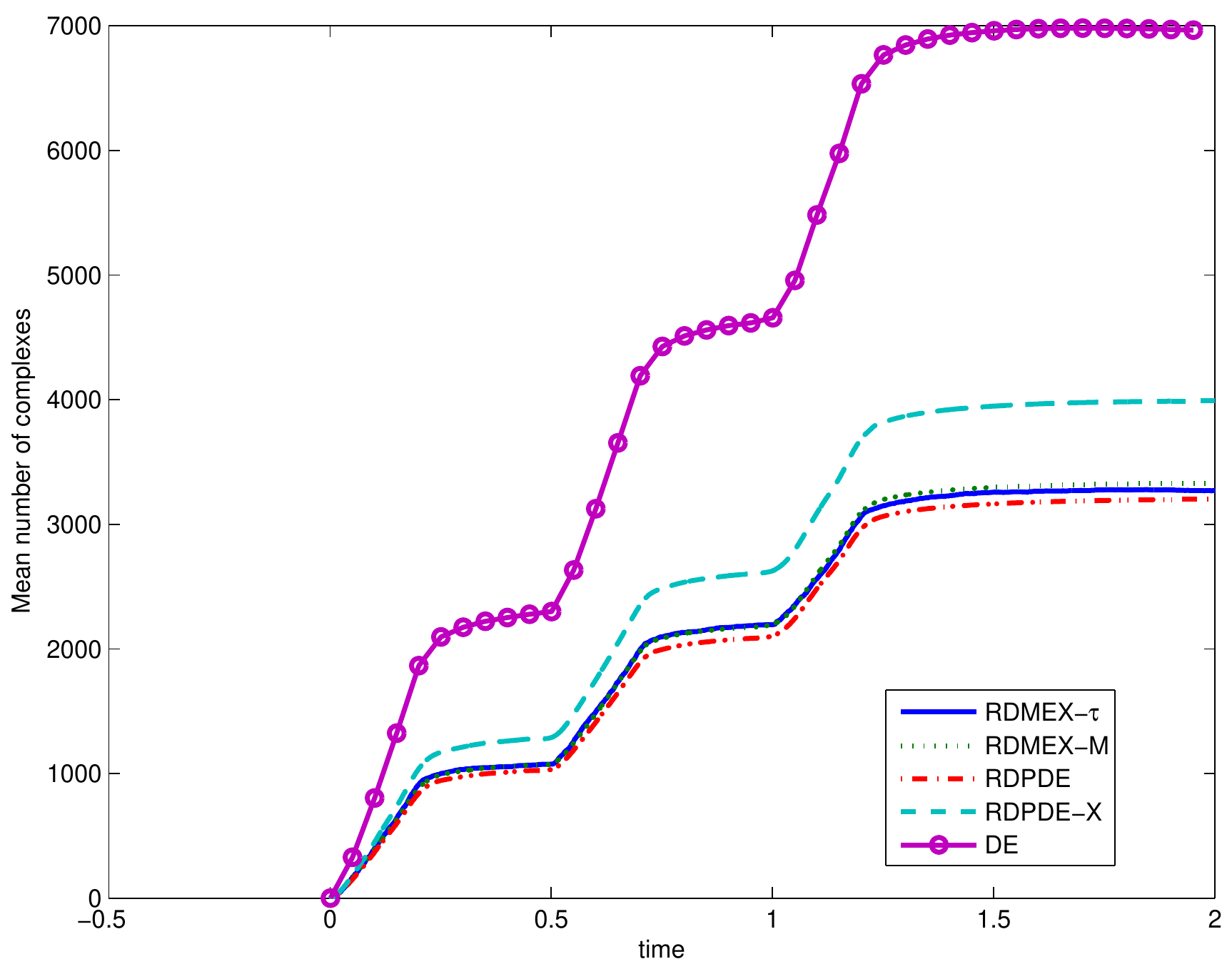}
\caption{The mean number of complexes in the 1-transmitter 1-receiver network for $K = 2.5\times10^{-3}$. (Section \ref{sec:1t1r_kp})}
\label{fig:fig11}
\end{center}
\end{figure}

We use two different values for $k_+$. Figure \ref{fig:fig11} shows the results for $k_+ = 2.5\times10^{-3}$. The figure compares the mean number of complexes formed in the time interval $[0,2]$. Results are obtained from RDMEX-$\tau$ (simulation) and RDPDE, RDMEX-M, RDPDE-X and DE (analytical models). It can be seen that both RDPDE and RDMEX-M (our analytical solutions) match RDMEX-$\tau$ well. The model RDPDE-X does not give good approximation because the voxel size is not a good approximation for the receptor size. Henceforth, we will not consider RDPDE-X further. The decoupled model DE does not match RDMEX-$\tau$ for this value of $k_+$. 

\begin{figure}
\begin{center}
\includegraphics[width=10cm]{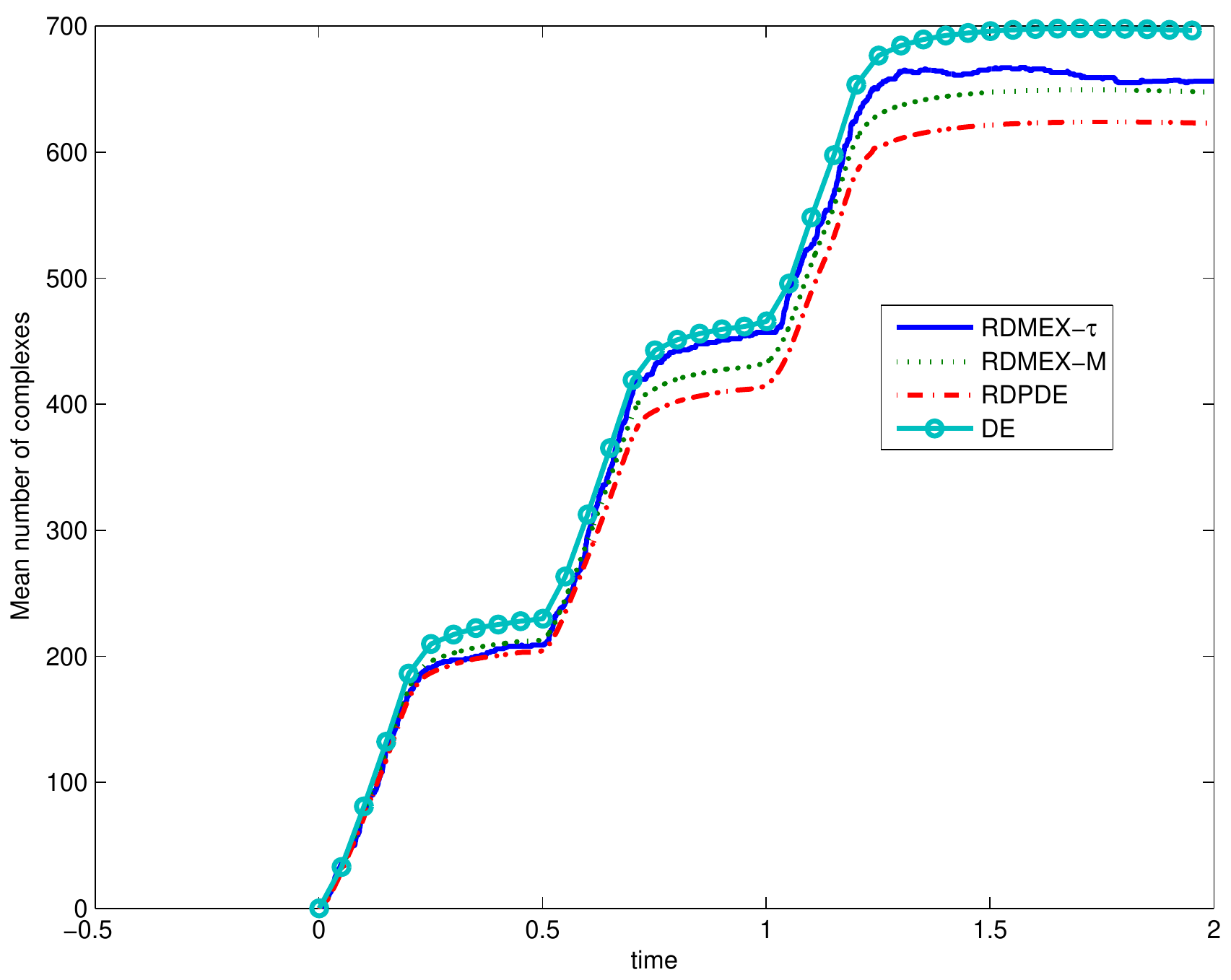}
\caption{The mean number of complexes in the 1-transmitter 1-receiver network for $K = 2.5\times10^{-4}$.(Section \ref{sec:1t1r_kp})}
\label{fig:fig12}
\end{center}
\end{figure}

We then change $k_+$ to $2.5\times10^{-4}$. The results are plotted in Figure \ref{fig:fig12}. Both RDMEX-M and RDPDE again match RDMEX-$\tau$ well. We also see that the decoupled model DE gives better prediction than before. This validates the discussion in Section \ref{sec:tf_int} that the decoupled model holds when $k_+$ is sufficiently small. 

\subsubsection{Accuracy of mean and standard deviation computation}
\label{sec:1t1r_cov}
In this experiment, we use a different transmitter signal to show the accuracy of RDMEX-M and RDPDE. We define two transmitter symbols $s_0$ and $s_1$. The symbol duration is 2 time units. When a transmitter sends $s_1$, it emits 10 molecules every $10^{-4}$ time units for a duration of 0.2 time units and then stops emitting for 1.8 time units. When a transmitter sends $s_0$, it does not emit any molecules for 2 time units. The transmitter signal $k_1(t)$ in this experiment is simply $s_1$. The receiver parameters are $k_+ = 2.5\times10^{-3}$ and $k_- = 8$. We determine the output signal $c_1(t)$ for $t \in [0,2]$. 

\begin{figure}
\begin{center}
\includegraphics[width=10cm]{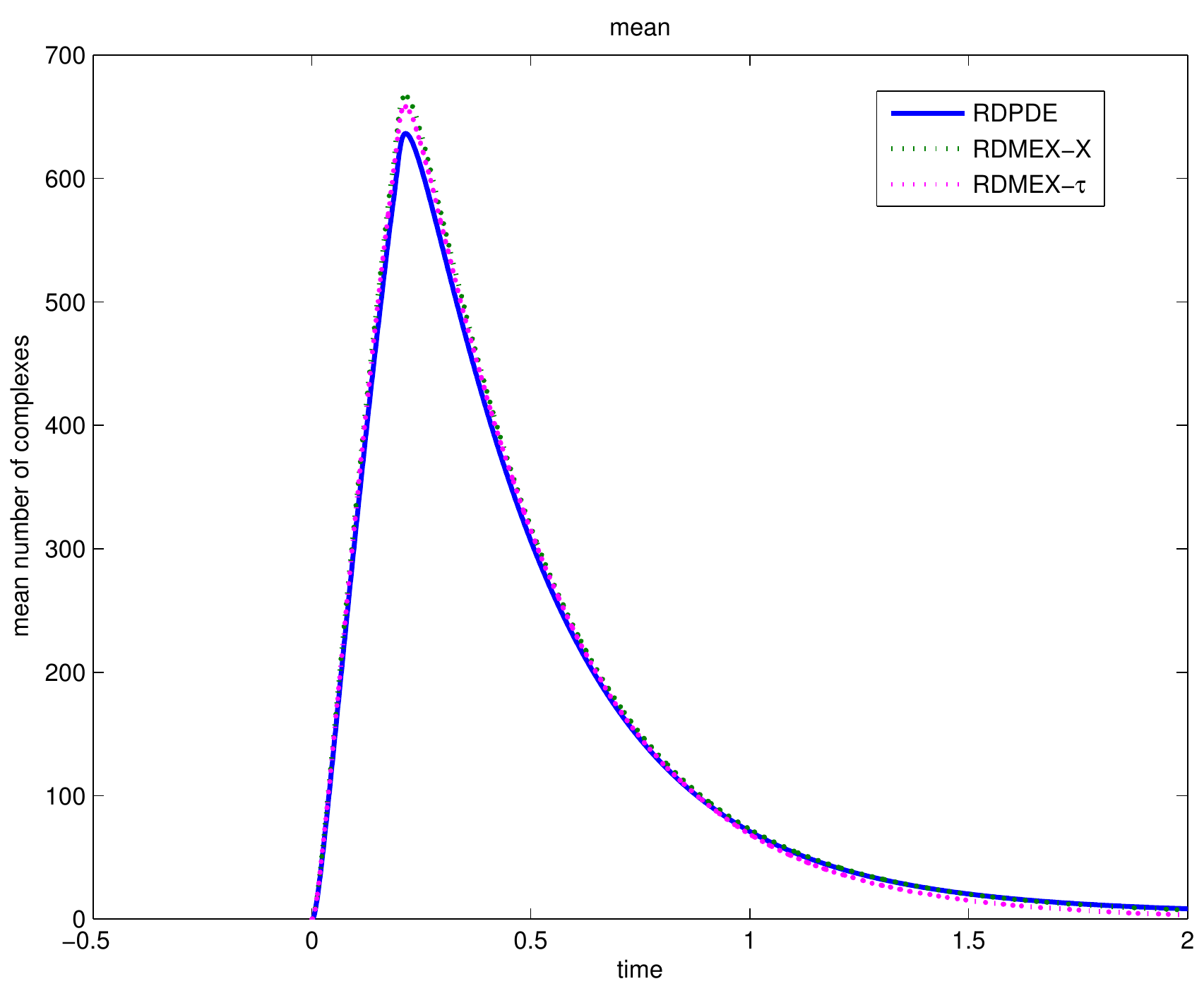}
\caption{The mean number of complexes in the 1-transmitter 1-receiver network with transmitter symbol $s_1$. (Section \ref{sec:1t1r_cov})}
\label{fig:fig11_s1_mean}
\end{center}
\end{figure}

We first verify the accuracy of using RDMEX-M and RDPDE to compute the mean of receiver output. For this experiment, we simulate the system using RDMEX-$\tau$ 125 times and compute the mean receiver output as the reference. Figure \ref{fig:fig11_s1_mean} shows the mean receiver output from RDMEX-M, RDPDE and RDMEX-$\tau$. It can be readily seen that RDMEX-M and RDPDE are accurate also for a different transmitter signal. 

\begin{figure}
\begin{center}
\includegraphics[width=10cm]{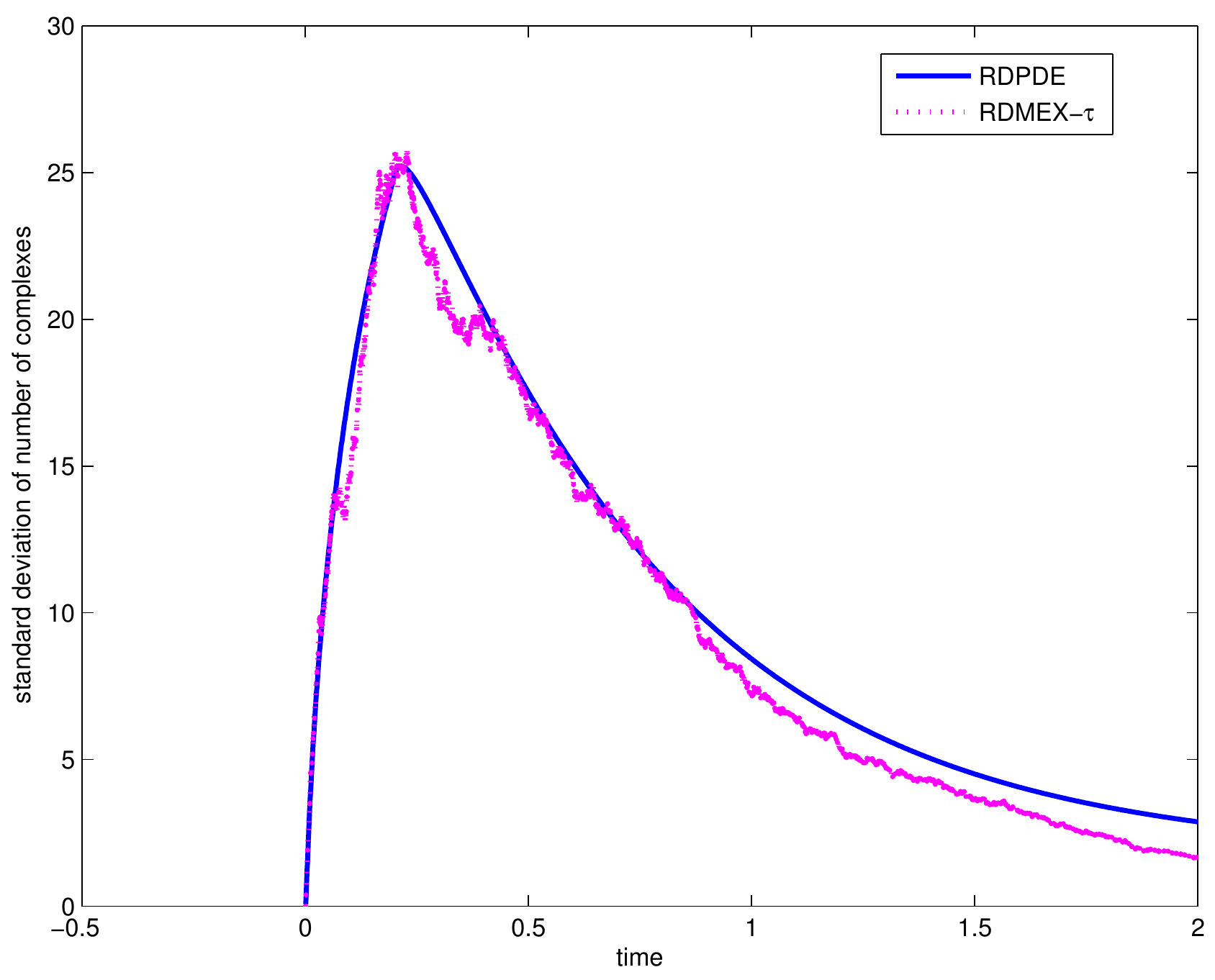}
\caption{The standard deviation of the number of complexes in the 1-transmitter 1-receiver network with transmitter symbol $s_1$. (Section \ref{sec:1t1r_cov})}
\label{fig:fig11_s1_cov}
\end{center}
\end{figure}

Our next goal is to verify the accuracy of using equation \eqref{eqn:rdmex_2m} to determine the standard deviation of the receiver output. We extend the RDMEX-M to solve for both the receiver output as well as the system states, which are the number of signalling molecules in the voxels. The system states are used as the input to \eqref{eqn:rdmex_2m} and numerical integration is used to solve for \eqref{eqn:rdmex_2m}. For verification, we simulate the system using RDMEX-$\tau$ 125 times to compute the standard deviation of the receiver output. Figure \ref{fig:fig11_s1_cov} plots the standard deviation of the receiver output from the two methods. It can be seen that \eqref{eqn:rdmex_2m} is accurate. 

One can envisage using the symbols $s_1$ and $s_0$ to encode the communication between the transmitter and the receiver. The communication scheme is similar to ON-OFF keying. For decoding, the receiver can use, say, the peak number of complexes to detect the symbol transmitted. If the peak number of complexes is above a threshold, then $s_1$ has been transmitted; otherwise, $s_0$ has been transmitted. 
 
\begin{remark}
The above method of calculating the variance using \eqref{eqn:rdmex_2m} is computational intensive. A future work is to derive efficient algorithm to solve  \eqref{eqn:rdmex_2m}. 
\end{remark} 

\subsubsection{Assumption of linear receiver model}
\label{sec:1t1r_mm}
In this paper, we assume that the rate of formation of complex $C$ is a linear function of the number of signalling molecules in the receiver voxel, according to the following chemical reaction kinetics: 
\begin{align}
\cee{L <=>[\text{k$_{1+}$}][\text{k$_{1-}$}]  C } \label{cr:sr}
\end{align} 

Under certain assumptions, the above reaction kinetics can be used to approximate more complex reactions. Consider the following chemical reactions: 
\begin{align}
\cee{L + E &<=>[\text{g$_{1+}$}][\text{g$_{1-}$}] I ->[\text{g$_{2}$}] C + E \label{cr:mmf} \\ 
C &->[\text{g$_{3}$}] L} \label{cr:mmr}
\end{align} 
Reaction \eqref{cr:mmf} is of Michaelis-Menten \cite{Jackson} type where a molecule $L$ reacts with an enzyme $E$ to form an intermediate product $I$, following by the decomposition $I$ into a product $C$ and the enzyme $E$. It can be shown that, for suitable choice of reaction constants in \eqref{cr:mmf}, the reaction kinetics of \eqref{cr:mmf} can be approximated by the forward reaction in \eqref{cr:sr}. Similarly, reaction \eqref{cr:mmr} can be made identical to the reverse reaction in \eqref{cr:sr} by choosing $g_3 = k_-$. 

\begin{figure}
\begin{center}
\includegraphics[width=10cm]{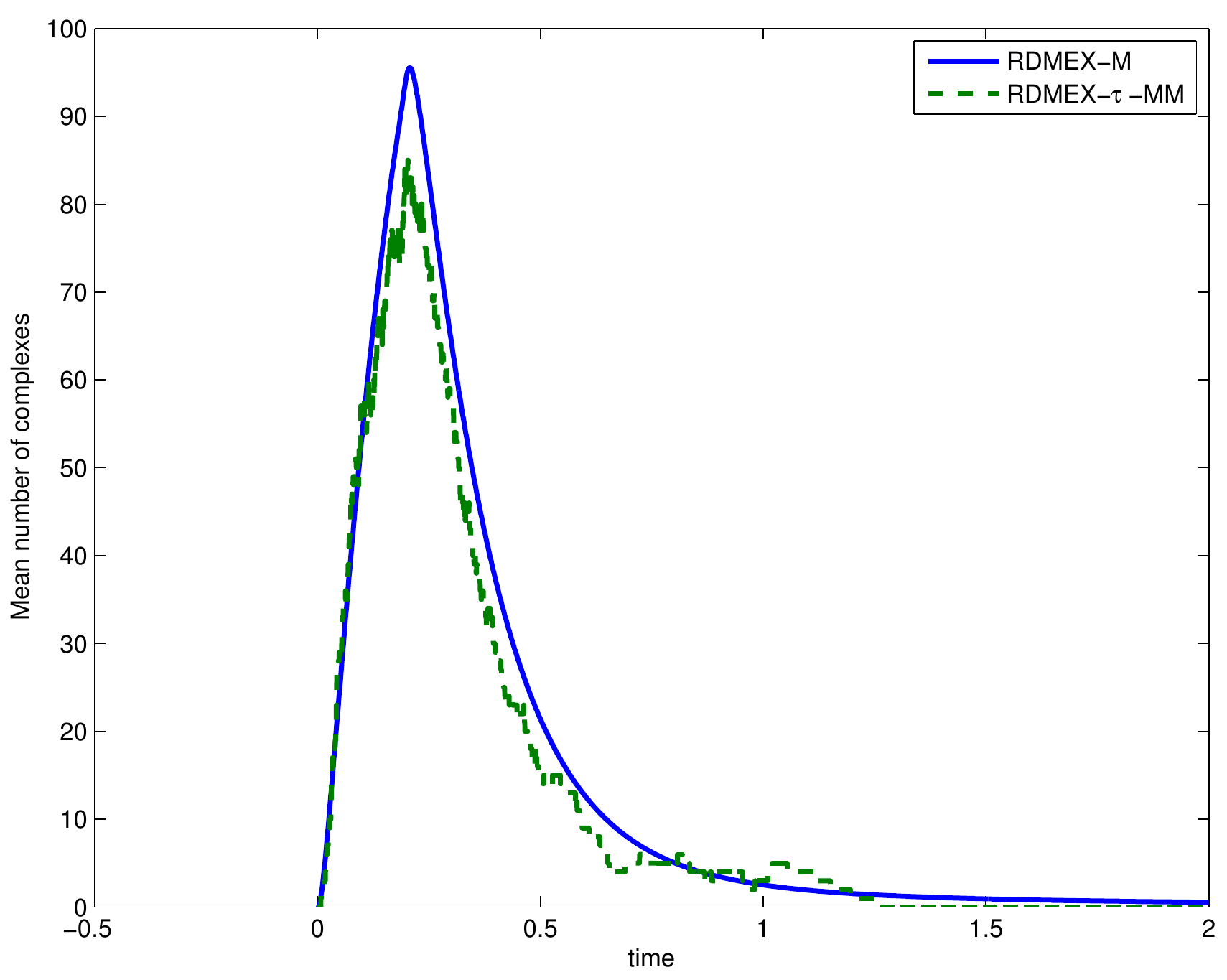}
\caption{The mean number of complexes in the 1-transmitter 1-receiver network for section \ref{sec:1t1r_mm}.}
\label{fig:fig11_mm}
\end{center}
\end{figure}

We want to show that a receiver with reactions \eqref{cr:mmf} and \eqref{cr:mmr} gives similar output signal compared to one with reactions \eqref{cr:sr}. We implement the receiver kinetics \eqref{cr:mmf} and \eqref{cr:mmr} in $\tau$-leaping simulation and refer to this method as RDMEX-$\tau$-MM. We compare this against RDMEX-M with identical parameters to those used in section \ref{sec:1t1r_cov}. The parameters in reactions \eqref{cr:mmf} and \eqref{cr:mmr} have been chosen to approximate those in \eqref{cr:sr}. The results are plotted in Figure \ref{fig:fig11_mm} and it can be seen that RDMEX-M is able to approximate more complicated reactions. 

\subsection{Effect of $\Delta$} 
\begin{figure}
\begin{center}
\includegraphics[width=10cm]{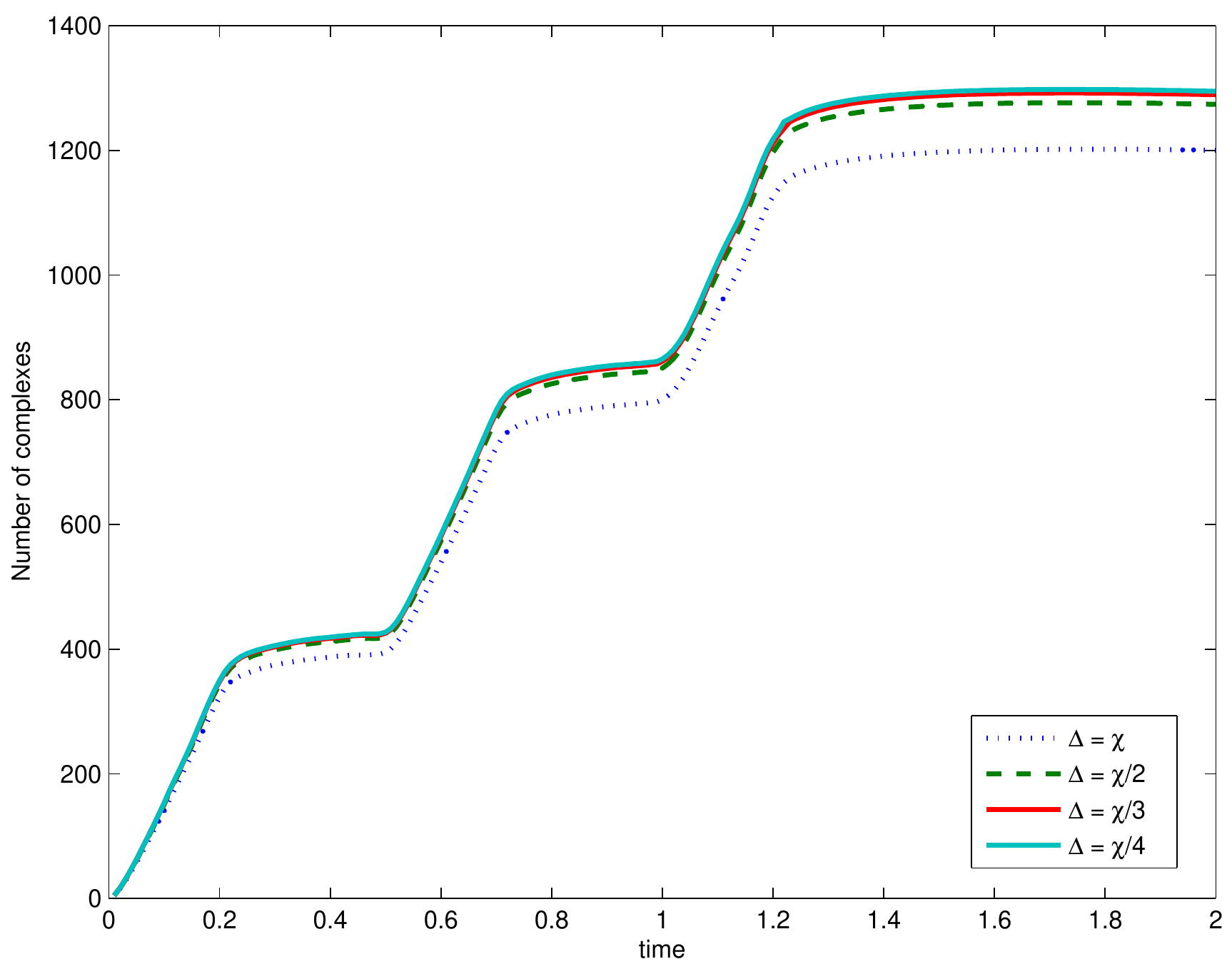}
\caption{This figure shows the effect of the size of a voxel on the mean number of complexes. The $\Delta$ values used are $\chi, \frac{\chi}{2}, \frac{\chi}{3}$ and $\frac{\chi}{4}$.}
\label{fig:delta}
\end{center}
\end{figure}

In this section, we study the effect of the size of voxels. We use the same parameter setting for the single-transmitter single-receiver case in Section \ref{sec:1t1r_kp}. We assume both the transmitter and receiver is a cube whose length of an edge is $\chi = 0.01$. We use four different voxel sizes with $\Delta = \chi, \frac{\chi}{2}, \frac{\chi}{3}$ and $\frac{\chi}{4}$. Since the size of the transmitter (or receiver) is a constant, this means that the transmitter (or receiver) occupies, respectively, 1, 8, 27 and 64 voxels for these 4 different voxel sizes. We assume that the emissions from the transmitter is uniformly distributed across all the voxels that it occupies. Figure \ref{fig:delta} show the mean number of complexes at the receiver, which is the sum of the number of complexes at all receiver voxels. It can be seem that the predicted output for $\Delta = \frac{\chi}{2}, \frac{\chi}{3}$ and $\frac{\chi}{4}$ are almost the same. This shows that as long as $\Delta$ is sufficiently small, the prediction is independent of $\Delta$.

\subsection{Single-transmitter two-receiver case}
\label{sec:1t2r} 
Equation \eqref{eqn:tf} shows that when there are multiple receivers, it is possible for a receiver to affect the output of another receiver. We will illustrate this phenomenon. We consider three different networks. Network 0 consists of a transmitter and two receivers, called 1 and 2. The transmitter, receivers 1 and 2 are located, respectively, at voxel $[0,0,0]$, $[1,0,0]$ and $[2,0,0]$. Network 1 is composed of the transmitter and receiver 1 of network 0. Network 2 consists of the transmitter and receiver 2 of network 0. For all the three networks, the transmitted signal is $s_1$ and the receiver parameters are $k_+ = 2.5\times10^{-3}$ and $k_- = 8$.

\begin{figure}
\begin{center}
\includegraphics[width=10cm]{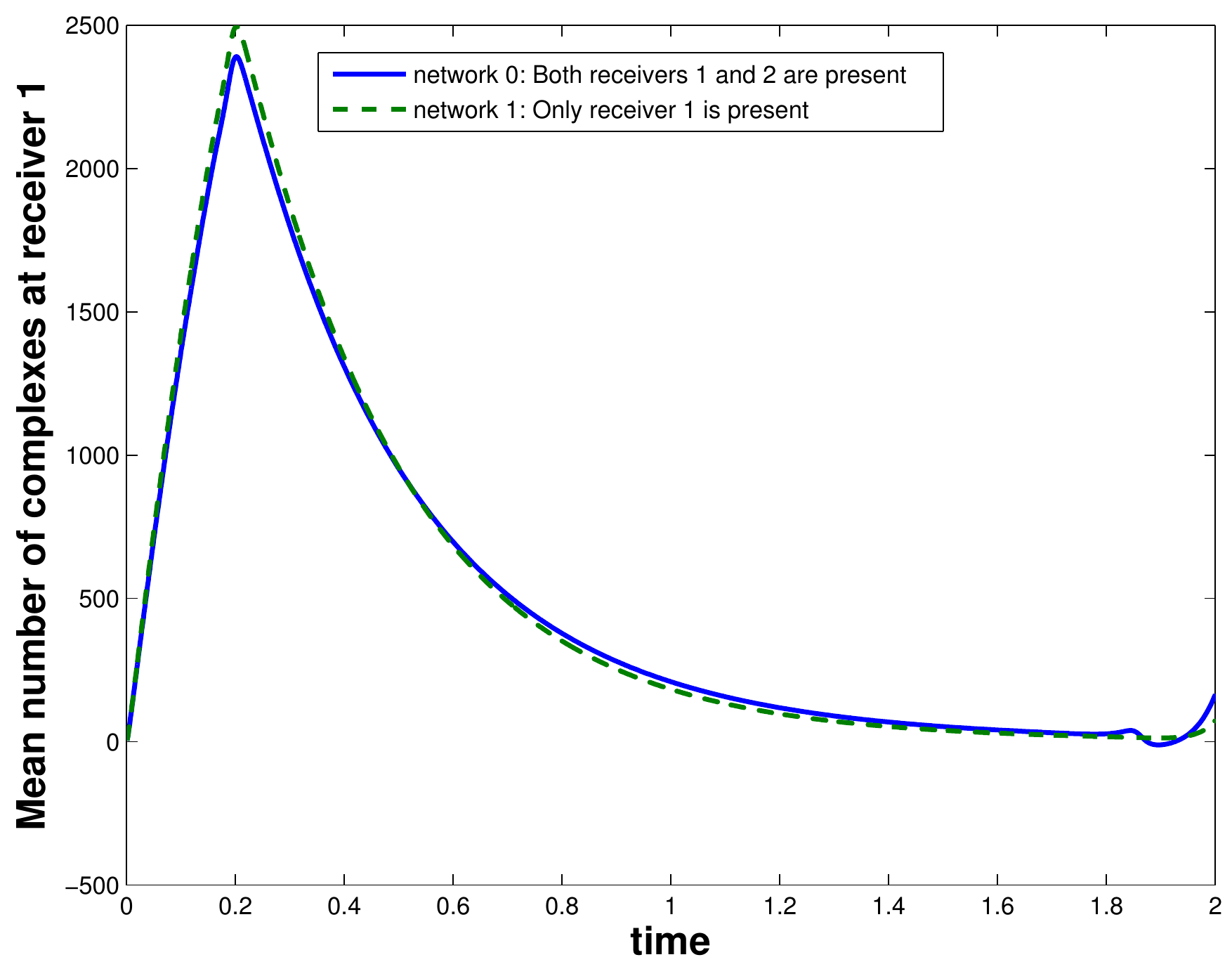}
\caption{This figure compares the output signal of two different networks. Network 0 consists of a transmitter and 2 receivers (receivers 1 and 2). Network 1 consists of the transmitter and receiver 1 of Network 0. The figure shows the output signal of receiver 1 for these two networks.}
\label{fig:fig12_fig1}
\end{center}
\end{figure}

\begin{figure}
\begin{center}
\includegraphics[width=10cm]{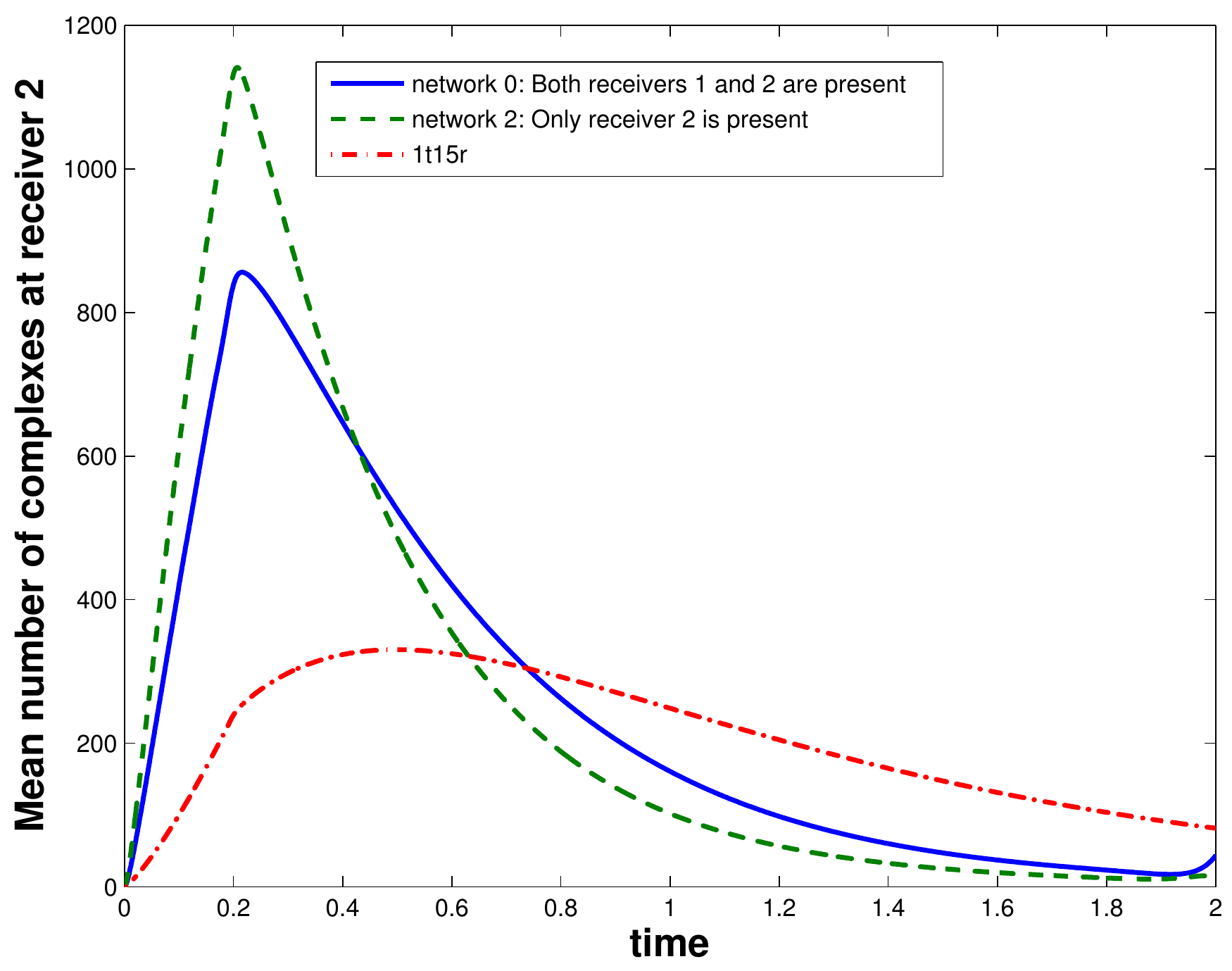}
\caption{This figure compares the output signal of three different networks. Network 0 consists of a transmitter and 2 receivers (receivers 1 and 2). Network 2 consists of the transmitter and receiver 2 of Network 0. The third network (label as 1t15r) consists of network 2 plus another 14 receivers. The figure shows the output signal of receiver 2 for these networks. }
\label{fig:fig12_fig2}
\end{center}
\end{figure}

Figure \ref{fig:fig12_fig1} shows the output for receiver 1 for networks 0 and 1, while Figure \ref{fig:fig12_fig2} shows the output for receiver 2 for networks 0 and 2. (The curve with labelled \verb|1t15r| in Figure \ref{fig:fig12_fig2} will be explained later.) It can be seen that the output of receiver 1 is almost the same for both network 0 (receiver 2 present) and network 1 (receiver 2 absent). However the output of receiver 2 for network 0 (receiver 1 present) is very different from that in network 2 (receiver 1 absent). Specifically, when receiver 1 is present, the output of receiver 2 has a lower peak number of complexes and a higher number of complexes at the tail. 

An explanation of how receiver 1 affects the output of receiver 2 is as follows. Note that receiver 1 is situated in between the transmitter and receiver 2. Some signalling molecules that reach receiver 2 have to pass through the voxel containing receiver 1. When these signalling molecules are in the receiver 1 voxel, some of them react to form complexes and are held up in the voxel. This means less signalling molecules reach receiver 2 in the early part of the symbol duration, thus resulting in a lower peak number of complexes in receiver 2. During the later part of the symbol duration, the complexes in receiver 1 dissociate to release the signalling molecules. Some of these signalling molecules, which are held up earlier in receiver 1, reach receiver 2 later on. This means more signalling molecules reach receiver 2 in the later part of the symbol duration if receiver 1 is present. This explains the behaviour at the tail of the output of receiver 2. 

When the number of receivers in a network is high, the output of some receivers in a network can be highly affected by the presence of the other receivers. We create a network with a large number of receivers by adding an additional 14 receivers to network 2, making a total of 15 receivers in the network. We again focus on the output of receiver 2. The curve labelled with \verb|1t15r| in Figure \ref{fig:fig12_fig2} shows the output of receiver 2 for this 1-transmitter 15-receiver network. It can be seen that the receiver output in this network is very different from that receiver 2 in network 2. The reason is that the other 14 receivers are affecting the output of receiver 2. 

The above results mean that the design of molecular communication need to take all receivers in the network into consideration. For example, if receiver 2 uses a threshold on peak number of complexes to detect $s_1$, we can see from Figure \ref{fig:fig12_fig2} that a threshold that works for the 1-transmitter 2-receiver network may not necessary work for the 1-transmitter 15-receiver network. 

\subsection{Two-transmitter Two-receiver case}
\label{sec:2t2r} 
We consider a molecular communication network with 2 transmitters located at $[0,0,0]$ (transmitter 1) and $[4,0,0]$ (transmitter 2), and 2 receivers located at $[1,1,0]$ (receiver 1) and $[2,0,1]$ (receiver 2). Both transmitters use $s_1$ and $s_0$ defined earlier as their transmission symbols. The signals for transmitters 1 and 2 are, respectively,  $s_0 s_1$ and $s_1 s_0$. 

\begin{figure}
\begin{center}
\includegraphics[width=10cm]{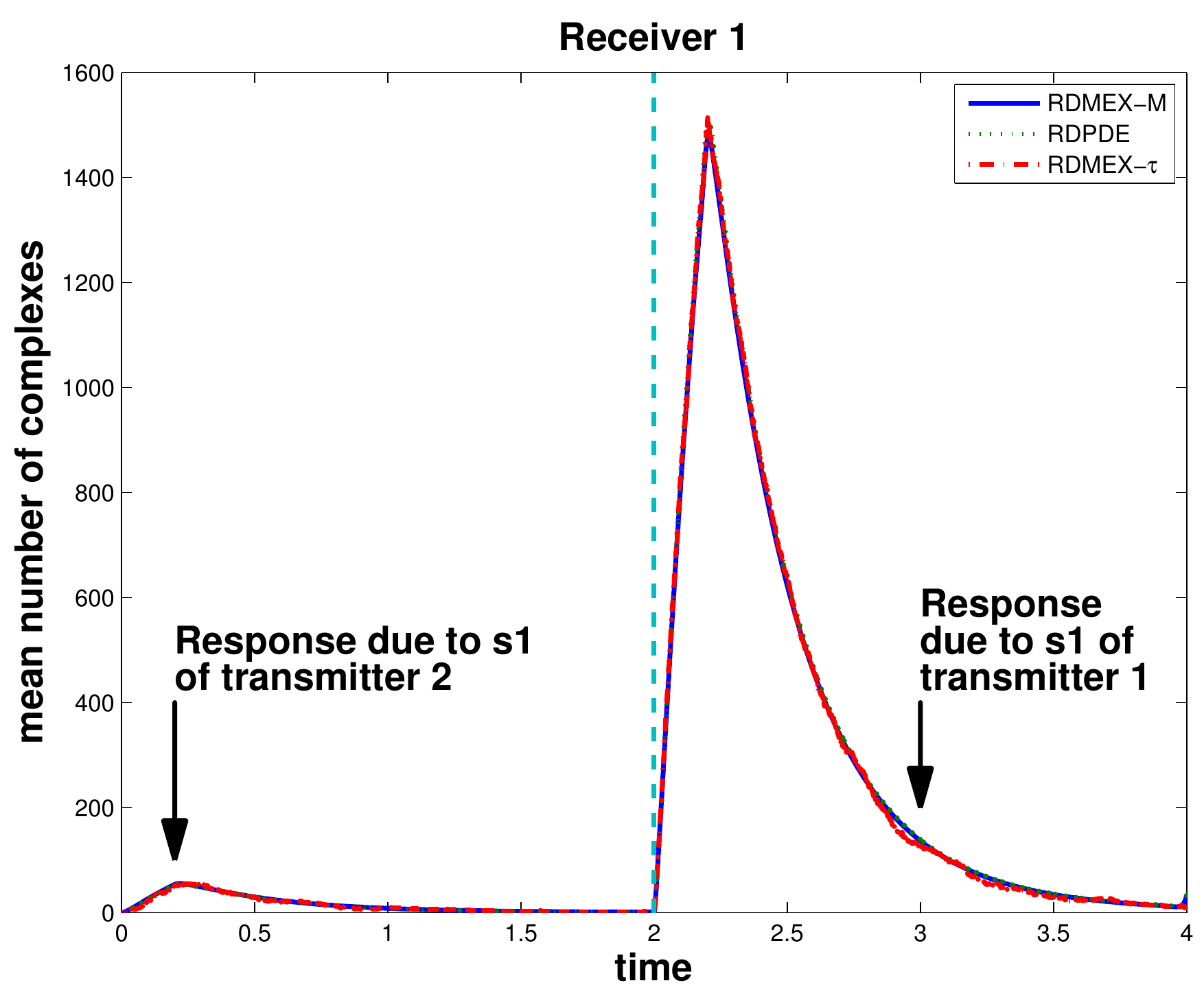}
\caption{The output signal of receiver 1 in the 2-transmitter 2-receiver network.}
\label{fig:fig21}
\end{center}
\end{figure}

\begin{figure}
\begin{center}
\includegraphics[width=10cm]{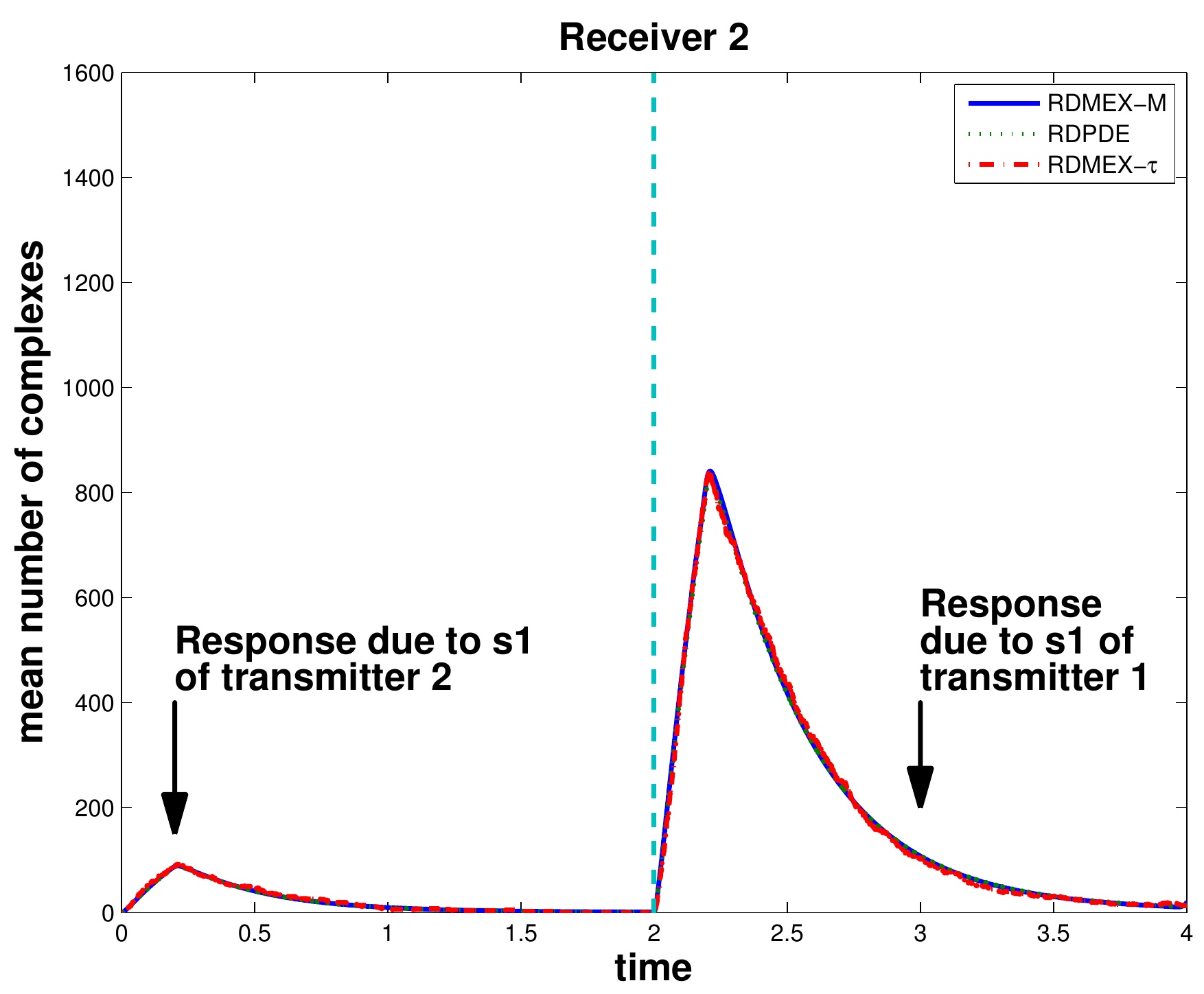}
\caption{The output signal of receiver 2 in the 2-transmitter 2-receiver network.}
\label{fig:fig22}
\end{center}
\end{figure}

We first verify the accuracy of our proposed models. Figures \ref{fig:fig21} and \ref{fig:fig22} show the output signals for, respectively, receivers 1 and 2. The three curves in each figure are obtained from RDMEX-$\tau$ (simulation) and RDPDE and RDMEX-M (our analytical solutions). It can be seen from both figures that the prediction from both RDPDE and RDMEX-M match that of RDMEX-$\tau$. In fact, the curves in the figures match so well that they overlap. 

We now assume the transmitters and receivers form two unicast communication pairs: transmitter 1 communicates with receiver 1 while transmitter 2 communicates with receiver 2. If transmitter 1 were the only transmitter, then receiver 1 should have a zero signal in the first symbol duration (between 0 and 2 time units). However, Figure \ref{fig:fig21} shows that receiver 1 has a non-zero signal during the first symbol duration due to the transmitter 2 sending an $s_1$ during this time. During the second symbol duration, the output signal of receiver 1 is due entirely to transmitter 1. If receiver 1 uses a threshold based detector, then a suitable choice of threshold will enable receiver 1 to correctly decode the two symbols sent by transmitter 1. 

Let us now consider the output signal of receiver 2 shown in Figure \ref{fig:fig22}. The signal in the first symbol duration is due to transmitter 2 (the intended signal) while that in the second symbol is due to transmitter 1 (the unintended signal). If receiver 2 uses a threshold based detector, then due to the unintended signal, a bit error will occur in the second symbol duration. This is a typical example of bit error when multiple transmitters and receivers share a common communication channel.

\section{Related work} 
\label{sec:related}
Molecular communication networks can be divided into two categories, according to whether they are natural or synthetic. Natural molecular communication networks are prevalent in living organisms. Their synthetic counterparts, though still rare, do exist. For example, \cite{Basu:2005cq} presents a system with multiple genetically engineered cells that use cell signalling to coordinate their behaviour. 
 
The modelling of natural and synthetic molecular communication networks is studied in different disciplines. The former is mainly studied in biophysics and mathematical physiology, while the latter in synthetic biology. There is also a recent interest in the engineering community to study molecular communication networks from a communication theory point of view \cite{Pierobon:2010vg,Atakan:2010bj,Srinivas:2010va}. This gives rise to a new research area called nano communication networks \cite{Akyildiz:2008vt}. 

Despite the fact that molecular communication networks are studied in diverse disciplines, the set of mathematical models that are being used are similar. This is not surprising given that the primary goal is to model diffusion and reaction kinetics. The classes of mathematical models being used include molecular dynamics, master equation, partial differential equation (PDE), Fokker-Planck equation, Langevin equation and others \cite{Gardiner}. We will focus on the first three classes of models in this discussion. 

Molecular dynamics is commonly used in simulation of molecular communication networks. Many examples of simulators exist, especially for natural molecular communication networks, see \cite{Burrage:2011te} for a recent overview. For synthetic networks, a recent example is \cite{Gul:2010es}. By analysing the molecular dynamics of transmitters and receivers, \cite{Pierobon:2011vr} characterises the noise in transmitters and receivers as, respectively, sampling and counting noise. 

There are ample examples in using PDE --- in particular diffusion PDE, telegraph equation and RDPDE --- to model molecular communication. For natural networks, \cite{Bialek:2005vq,Endres:2009vx} use RDPDE to study the noise in receptor binding in chemotaxis, and \cite{Neufeld:2009ug} uses RDPDE to study signalling cascades. However, these papers do not consider the transmitters. For synthetic networks, telegraph or diffusion PDEs (or their kernels) have been used to characterise the diffusion of signalling molecules in \cite{Pierobon:2011ve,Mahfuz:2011te,Arifler:2010tba,Atakan:2010bj,Srinivas:2010va} and others. However, these papers do not consider the coupling effect between diffusion and receiver reaction kinetics. In our earlier work in \cite{Chou:nano_j}, we use a RDPDE in the form of \eqref{eqn:rdpde_3d}, as a deterministic model for molecular communication network. The RDPDE in \cite{Chou:nano_j} is solved numerically and no analytic solution is provided. In this paper, we derived a RDPDE model \eqref{eqn:rdpde_3d} for molecular communication and provide an interpretation of the model as the mean receiver output of molecular communication networks. In addition, we present an analytical solution to this RDPDE and show that it can be used to accurately predict mean receiver output in molecular communication networks. 

For some time, RDME has been considered to be a phenomenological model because it diverges in certain cases \cite{Isaacson:2008jc}. Fortunately, the problem has been resolved in \cite{Fange:2010wk} and there is now a firm theoretic basis for RDME. There are many examples of work that use RDME to model natural molecular communication networks, see \cite{Isaacson:2007ui,Elf:2004ud}. However, these papers do not consider the transmitters. The use of RDME in studying synthetic molecular communication networks appear to be novel. To the best of our knowledge, our RDMEX model, which is formed by coupling time sequences of signalling molecule emission pattern with RDME, has not been proposed before. The proposed RDMEX model is one of the novel contributions of this paper. 

\section{Conclusions and future work} 
\label{sec:con}
In this paper, we have proposed a new stochastic model called reaction-diffusion master equation with exogenous input (RDMEX) for modelling molecular communication networks with multiple transmitters and multiple receivers. We show that we can readily derive the mean and covariance of receiver output of RDMEX model for the case where reaction kinetics at the receiver is linear. We then solve the mean receiver output of RDMEX model using two different methods. In the first method, we derive the continuum limit of RDMEX and present a closed-form expression of the solution to the resulting reaction-diffusion partial differential equation. In the second method, we solve the mean receiver output of RDMEX explicitly by using transform techniques. We present numerical examples comparing the accuracy of our analytical solutions against simulation. We find our analytical solutions give accurate prediction of mean receiver output of molecular communication networks. This paper has focused on studying the mean receiver output of molecular communication networks. However, one needs to understand the properties of noise in these networks in order to evaluate their performance, such as bit error rate or capacity. Future work includes the study of the properties of noise in the RDMEX model and using the RDMEX model to evaluate the performance of molecular communication networks. 

\appendix
\section{Derivation of \eqref{eqn:Ltilde}}
\label{app:rdpde}
Equation \eqref{eqn:Ltilde} can be derived by using a couple of different methods, e.g. Fourier transform or Green's function. We will use Green's function here. 

Note that the RDPDE \eqref{eqn:rdpde_3d} is an inhomogeneous partial differential equation where the last two terms act as the forcing function. Let $G(v,t)$ denote the three dimensional kernel of the linear diffusion equation $\frac{\partial \ell }{\partial t} = D \nabla^2 \ell$. Let also $\ast_{S,T}$ denote convolution in space and time, and $\ast_{T}$ denote convolution in time only. By using the theory of Green's function, we have 

\begin{eqnarray}
\ell(v,t) & = & G(v,t) \ast_{S,T} \left\{ \sum_{a = 1}^2 \delta(v - v_{T,a}) k_a(t)  - \sum_{u = 1}^2 \delta(v - v_{R,u}) \frac{d c_u(t)}{dt} \right\} \nonumber \\
& = & \sum_{a = 1}^2 G(v - v_{T,a},t) \ast_{T} k_a(t)  - \sum_{u = 1}^2 G(v - v_{R,u},t) \ast_{T}  \frac{d c_u(t)}{dt} 
\label{eqn:app:ell} 
\end{eqnarray}

Now we can obtain \eqref{eqn:Ltilde} from applying temporal Fourier transform to \eqref{eqn:app:ell}, noting that $\phi(v,\omega)$ is the temporal Fourier transform of $G(v,t)$. 

\section{Derivation of \eqref{eqn:Ltilde_d}}
\label{app:z} 
Note that \eqref{eqn:mean_3d_1} is a linear difference equation in space and a continuous differential equation in time. We can draw a parallel with the theory of Green's function and note that \eqref{eqn:Ltilde_d} holds if $\psi(\xi,\omega)$ is the temporal Fourier transform of the solution of the following equation:
\begin{eqnarray}
\frac{d h_{i,j,k}(t) }{dt} & = & {D}  (h_{i-1,j,k}(t)  - 2 h_{i,j,k}(t) + h_{i+1,j,k}(t) ) + 
{D} (h_{i,j-1,k}(t)  - 2 h_{i,j,k}(t) + h_{i,j+1,k}(t) ) + \nonumber \\ 
& & {D} (h_{i,j,k-1}(t) - 2 h_{i,j,k}(t) + h_{i,j,k+1}(t) ) + \frac{1}{\Delta^3} \delta_K(\xi) \delta(t)  
\label{eqn:diff_cont}
\end{eqnarray} 
This equation can be solved by using Fourier transform on the continuous variable $t$ and z-transform on the discrete variables $i$, $j$ and $k$. Let $H_{i,j,k}(\omega)$ denote the Fourier transform of $h_{i,j,k}(t)$. The multi-dimensional z-transform of $H_{i,j,k}(\omega)$ is defined as: 
\begin{eqnarray}
\Psi(W_x,W_y,W_z,\omega) & = & \sum_{i = -\infty}^{\infty} \sum_{j = -\infty}^{\infty} \sum_{k = -\infty}^{\infty} 
H_{i,j,k}(\omega) W_x^{-i} W_y^{-j} W_z^{-k} 
\end{eqnarray}
By using \eqref{eqn:diff_cont}, it can be shown that 
\begin{eqnarray}
\Psi(W_x,W_y,W_z,\omega) & = & 
\frac{-1}{{D} \Delta \{(W_x - W_x^{-1})^2 + (W_y - W_y^{-1})^2 + (W_z - W_z^{-1})^2  - \frac{\iota \omega \Delta^2}{\tilde{D}} \} }
\label{eqn:app:Psi} 
\end{eqnarray}
We first compute inverse z-transform with respect to $W_z$. We re-write \eqref{eqn:app:Psi} as 
\begin{eqnarray}
\Psi(W_x,W_y,W_z,\omega) & = & 
\frac{-1}{{D} \Delta} 
\frac{W_z}{(W_z - W_{z\ast})(W_z - W_{z\ast}^{-1}) }
\end{eqnarray}
where $W_{z\ast}$, as defined in section \ref{sec:zt}, is the root of \eqref{eqn:qe} with modulus less than unity. Note we have also used the fact that the roots of \eqref{eqn:qe} are reciprocal of each other. It can now be shown that the inverse z-transform of $\Psi(W_x,W_y,W_z,\omega)$ with respect to $W_z$ is 
\begin{eqnarray}
\frac{-1}{{D} \Delta} \frac{W_{z\ast}^{|k|+1}}{W_{z\ast}^2-1} 
\end{eqnarray}
If we apply the standard inverse z-transform contour integrals, with respect to $W_x$ and $W_y$, to this expression, then we obtain the formula for $\psi(v,\omega)$ given in \eqref{eqn:psi}. 

\bibliographystyle{IEEETran}
\bibliography{nano,book}

\newpage
\listoffigures 

\end{document}